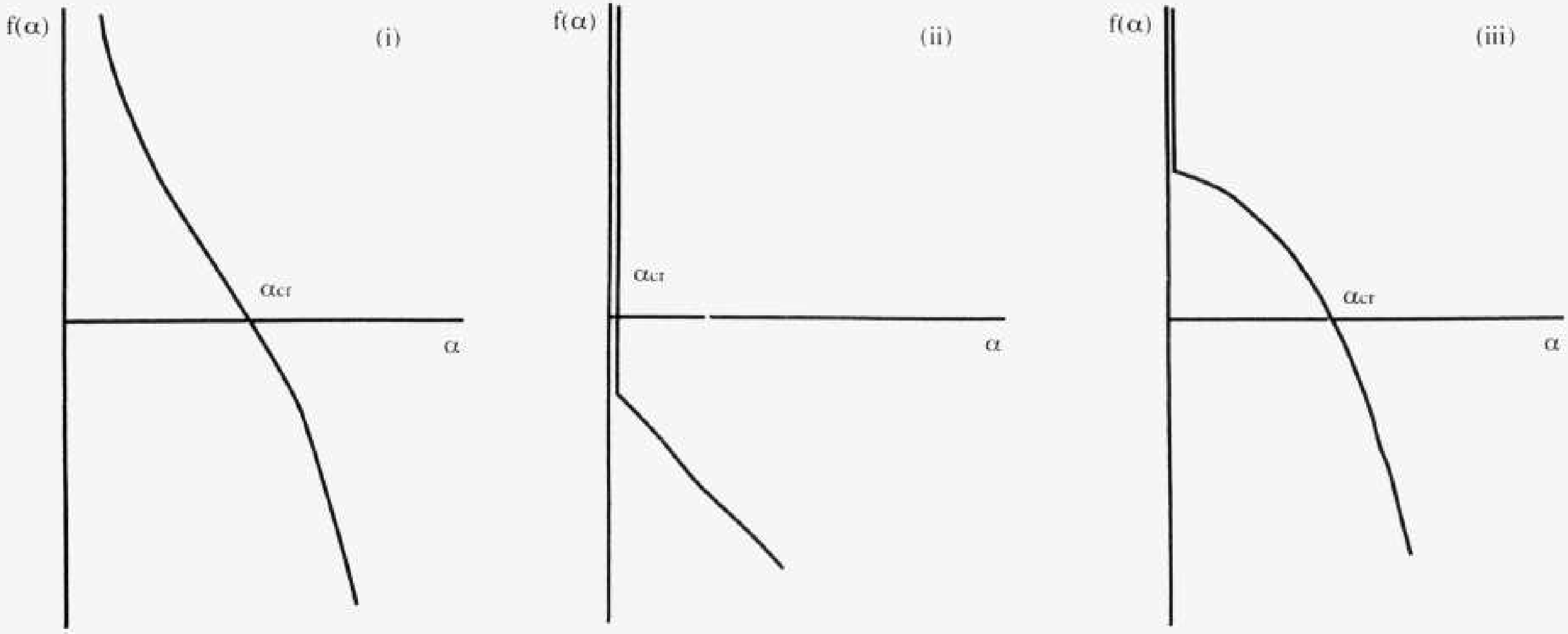

Figure 1

# LATTICE COULOMB HAMILTONIAN AND STATIC COLOR-COULOMB FIELD

DANIEL ZWANZIGER†

*Physics Department, New York University*

*New York, N. Y. 10003*

ABSTRACT

The lattice Coulomb-gauge hamiltonian is derived from the transfer matrix of Wilson's Euclidean lattice gauge theory, wherein the lattice form of Gauss's law is satisfied identically. The restriction to a fundamental modular region (no Gribov copies) is implemented in an effective hamiltonian by the addition of a "horizon function" G to the lattice Coulomb-gauge hamiltonian. Its coefficient $\gamma_0$ is a thermodynamic parameter that ultimately sets the scale for hadronic mass, and which is related to the bare coupling constant $g_0$ by a "horizon condition". This condition determines the low momentum behavior of the (ghost) propagator that transmits the instantaneous longitudinal color-electric field, and thereby provides for a confinement-like feature in leading order in a new weak-coupling expansion.

---

† Research supported in part by the National Science Foundation under grant no. PHY93-18781

## 1. Introduction

Quantum chromodynamics, which appears to provide a satisfactory framework for understanding the strong interactions, is expected to have two striking features: (i) the force between colored particles is of such long range that it confines them, and so (ii) the gluons whose exchange is responsible for this force are absent from the physical spectrum. These two features, though complimentary, appear difficult to accommodate simultaneously in a manifestly Lorentz-covariant theory where one might naively expect that the gluon propagator at zero momentum should (i) possess a strong singularity as a manifestation of the long-range force and (ii) be strongly suppressed because there are no massless gluons.* By contrast, these two features are compatible in the Coulomb gauge because the long-range force can be carried by an instantaneous static color-Coulomb field, while the propagator of (three-dimensionally) transverse, would-be physical gluons is suppressed. The instantaneous field is required by Gauss's law, which is satisfied identically by the color-electric fields in the Coulomb-gauge hamiltonian. The restriction to the fundamental modular region, a region without Gribov copies [1], makes the Gauss's law constraint more severe. Indeed many years ago Gribov [1] proposed this as a confining mechanism.

* The ultimate consistency of a Lorentz or Euclidean-lattice covariant gauge is not in question. However the Coulomb gauge presents a simpler *picture* of confinement and allows more extensive calculations than the author has so far achieved in a covariant gauge despite considerable effort.

We derive the lattice Coulomb-gauge hamiltonian by gauge-fixing in Wilson's Euclidean lattice gauge theory. This is a quantized, regularized and gauge-invariant theory which one is free to gauge fix. After gauge-fixing, the lattice Coulomb-gauge hamiltonian is obtained from the transfer matrix of the Euclidean lattice theory by integrating out the variables on the vertical links in the limit in which the length of these links approaches zero. One finds that the lattice form of Gauss's law is satisfied exactly, and an instantaneous long-range confining potential appears in the lattice Coulomb-gauge hamiltonian. The gauge-fixed theory inherits the non-perturbative regularization of Wilson's theory, so ultra-violet and infra-red divergences are strictly controlled. Although cancellation of the ultraviolet divergences in the lattice Coulomb gauge is not yet established explicitly, this should be a technical problem as long as the cancellation occurs in Wilson's Euclidean gauge theory.

In classical Yang-Mills theory, Gauss's law in the presence of quark color-charge density $\rho_{qu}$ reads

$$(D\cdot E)^a \equiv \nabla\cdot E^a + g_0 f^{abc} A^b\cdot E^c = g_0\, \rho_{qu}{}^a, \qquad (1.1)$$

where A represents the 3 spatial components of the vector potential, and D is the 3-dimensional gauge-covariant derivative. (Vectors are always 3-vectors.) Gauss's law fixes the longitudinal component of the color-electric field. Indeed, if we decompose E into its transverse and longitudinal parts,

$$E_i^a(x) = E_i^{tr,a}(x) - \nabla_i \varphi^a(x), \qquad (1.2)$$

where $\nabla \cdot E^{tr} = 0$, and $\varphi$ is a color-Coulomb potential, then Gauss's law reads

$$- (D \cdot \nabla \varphi)^a = - g_0 f^{abc} A^b \cdot E^{tr,c} + g_0 \rho_{qu}^a \equiv g_0 \rho^a. \qquad (1.3)$$

Here a color-charge density $\rho^a$ is defined in which only $E^{tr}$ appears instead of E. The operator on the left – where one would find $-\nabla^2$ in an Abelian theory – is the 3-dimensional Faddeev-Popov operator $M(A) \equiv - D(A) \cdot \nabla$. Gauss's law determines the color-Coulomb potential $\varphi$ in terms of the remaining variables by means of the instantaneous and Faddeev-Popov propagator $M^{-1}(A)$,

$$\varphi^a(x) = g_0 (M^{-1} \rho)^a(x) = g_0 \int d^3y \, (M^{-1})^{a,b}(x, y; A) \, \rho^b(y). \qquad (1.4)$$

Despite its non-invariant appearance, the color-Coulomb law holds in every gauge and in every Lorentz frame. The color-electric field

$$E = E^{tr} - g_0 \nabla M^{-1} \rho, \qquad (1.5)$$

$$E^a = E^{tr,a} - g_0 \nabla (M^{-1})^{ab} (- f^{bcd} A^c \cdot E^{tr,d} + \rho_{qu}^b) \qquad (1.6)$$

is expressed in terms of the transverse color-electric field and (the spatial components of) A,

In the Coulomb gauge, one fixes $A_i^a(x) = A_i^{tr,a}(x)$. This makes $M(A) = M(A^{tr})$ symmetric. It also fixes $A^0 = - A_0$ by a time-independent constraint. For with $E = -\partial A/\partial t - DA^0$, one has $\nabla\cdot E = MA^0 = -\nabla^2\varphi$, so

$$A^0 = M^{-1}(-\nabla^2)\varphi = g_0 M^{-1}(-\nabla^2)M^{-1}\rho. \tag{1.7}$$

In the hamiltonian formulation in the Coulomb gauge, the conjugate canonical dynamical variables are $A^{tr}$ and $-E^{tr}$. The expressions that appear on the right-hand side of the last four equations are functions of these dynamical variables only. In the quantum theory, one sets $E^{tr} = i\partial/\partial A^{tr}$, and the continuum quantum Coulomb hamiltonian of Christ and Lee [2], is given by

$$H = 2^{-1} \int d^3x \, (\sigma^{-1}E^{\dagger}\sigma E + B^2), \tag{1.8}$$

Here E is as above, $\varepsilon_{ijk}B_k^a = \nabla_i A_j^{tr,a} - \nabla_j A_i^{tr,a} + g_0 f^{abc} A_i^{tr,b} A_j^{tr,c}$, and $\sigma \equiv \det M(A^{tr})$. The energy of the color-electric field (1.6) includes the instantaneous longitudinal field produced by the A-dependent integral operator $M^{-1}(A^{tr})$. We shall see that the restriction to the fundamental modular region makes this a confining field.

The lattice Coulomb hamiltonian we shall obtain is of canonical form and contains color-electric fields that satisfy the lattice form of Gauss's law identically. Indeed, in the coordinate system which we shall use, which on each spatial link is,

$$t^a A_{x,i}^a = 2^{-1}(U_{x,i} - U_{x,i}^\dagger)_{\text{traceless}}, \tag{1.9}$$

equations (1.2) – (1.6) are unchanged in form in going from the continuum theory to the lattice. (Here $U_{x,i}$ is Wilson's link variable that transforms under gauge transformation according to $U_{x,i} \to g_x^\dagger U_{x,i} g_y$, where $y = x+e_i$. To avoid possible confusion we emphasize that gauge-fixing is done after Wilson's gauge-invariant lattice quantization by compact variables.) The only change in formula (1.8) for the continuum quantum hamiltonian is that in the lattice Coulomb hamiltonian, the link variable $E_{x,i,a}$ is replaced by

$$\mathcal{E}_{x,i,a} = J_a{}^b(A_{x,i}{}^{tr})E_{x,i,b}, \tag{1.10}$$

where the $J_a{}^b(A)$ satisfy the differential equations of the Lie algebra of SU(n) on each link, namely $[J_a{}^c(A)\partial/\partial A^c, J_b{}^d(A)\partial/\partial A^d] = f^{abc}J_c{}^d(A)\partial/\partial A^d$ (and of course the magnetic energy is the plaquette action on spatial plaquettes, and the weight $\sigma$ includes the Haar measure). The lattice form of equations (1.4) – (1.7), and the Heisenberg equations of motion of the lattice Coulomb hamiltonian provide a quantized and regularized expression for all the classical Yang-Mills equations of motion. Use of the coordinates (1.9) instead of the exponential mapping $U_{x,i} = \exp(t^a A_{x,i}{}^a)$, may be essential to obtaining the lattice Coulomb hamiltonian in closed form, which otherwise appears to be known only as an infinite power series [3].

In the quantum gauge theory, in addition to the lattice Coulomb hamiltonian, one specifies a configuration space which is free of Gribov copies. (For an alternative approach, see [4] and [5].) This is the fundamental modular region described in sects. 3 and 5. Section 6 contains the crucial observation that the restriction to the fundamental modular region contradicts the usual perturbative expansion in powers of $g_0$. We find however that it is consistent with an expansion in powers of $g_0$, in which the leading contribution to $\langle M^{-1} \rangle$ is of order $g_0^{-1}$,

$$\langle (M^{-1})^{a,b}(x, y; A^{tr}) \rangle = g_0^{-1} u(x-y) \delta^{a,b} + O(1). \tag{1.11a}$$

This gives an x-dependent mean-field contribution to the inverse Faddeev-Popov matrix itself

$$(M^{-1})^{a,b}(x, y; A^{tr}) = g_0^{-1} u(x-y) \delta^{a,b} + O(1). \tag{1.11b}$$

Thus, by (1.5), the color-electric field acquires a longitudinal part that is of order $g_0^0$ instead of $g_0^1$,

$$E = E^{tr} - \nabla \int d^3y \, u(x - y) \rho(y) + O(g_0) \tag{1.12}$$

which contributes to the zero-order hamiltonian. This is a rather literal expression of anti-screening. We call u(r), which is the leading-order mean-field contribution to $g_0 M^{-1}$, the static color-Coulomb potential. It is independent of $g_0$ and $A^{tr}$, and is determined by a consistency condition, eq. (6.22) below.

The organization of this article is as follows. Section 2 is didactic in style, although the results are believed to be new. It addresses the problem of how a restriction on the configuration space, such as the restriction to a fundamental modular region $\Lambda$, may be implemented for a system with a large number of degrees of freedom, such as a lattice theory described by a hamiltonian H. The implementation relies on the close connection between a quantum mechanical hamiltonian and a classical statistical mechanical system with partition function $Z = \mathrm{tr}[\exp(-\beta H)] = \lim_{M\to\infty} T^M$, where the transfer matrix is given by $T = \exp(-\varepsilon H)$. Suppose that the configuration space is restricted by a condition of the form $G(A) \leq 0$, where $G(A)$ is a bulk quantity. In sect 2 it is shown that in the partition function the restriction may be implemented by using the transfer matrix $T_{eff} = \exp[ -\varepsilon\ (H+\gamma G)]$, and integrating over the unrestricted configuration space. Here $\gamma$ is a thermodynamic parameter that is determined in the infinite-volume limit by the condition

$$\lim_{V\to\infty} V^{-1}\langle G\rangle = 0_-, \tag{1.13}$$

where the expectation value is calculated in the system with partition function $Z_{eff} = \lim_{M\to\infty} T_{eff}{}^M$. Because of the equivalence between a statistical mechanical system and a quantum mechanical system, this corresponds to using the effective hamiltonian,

$$H_{eff} = H + \gamma G, \tag{1.14}$$

in the unrestricted configuration space, and calculating $\langle G\rangle$ in the ground state of $H_{eff}$. Thus the familiar equivalence of different statistical mechanical ensembles, such the micro-canonical and canonical, implies the equivalence of the corresponding quantum-mechanical systems. We call G, the horizon function, and (1.13) the horizon condition. The method is illustrated by some elementary examples. In particular, the eigenvalues of the N-dimensional laplacian operator restricted to the interior of an N-dimensional ellipsoid are calculated in the limit $N \to \infty$. The implementation of the restriction of the configuration space by (1.13) and (1.14) is correct only in the infinite-volume limit, where the number of degrees of freedom becomes infinite. On the contrary, in the works of Lüscher [6], Cutkosky [7], and van Baal [8] and co-workers, which also use the Coulomb hamiltonian, the restriction to the fundamental modular region is imposed in the approximation in which a finite number of modes are retained.

For the case of QCD, it will be found that the thermodynamic parameter $\gamma$ sets the scale for hadronic mass, and corresponds in conventional perturbation theory to the introduction of a renormalization mass $\Lambda_{QCD}$. The horizon condition (1.13) may be solved for $g_0$ instead of $\gamma$, which corresponds in conventional perturbation theory to solving the renormalization-group equation for $g_0$.

In sect. 3 we give a precise definition of the minimal lattice Coulomb gauge, which is used to fix Wilson's Euclidean gauge-invariant lattice theory. It is has the defining property that the link variables U are made as close to unity as possible, in an optimal way, on all the spatial links of the Euclidean lattice. Some elementary properties of this gauge are derived in sect. 3.

In sect. 4 and Appendix A, the lattice Coulomb hamiltonian $H_{coul}$, eq. (4.16), is derived from the transfer matrix associated with the partition function of Wilson's Euclidean lattice gauge theory,

$$Z = \int dU \exp[ - S_W(U) ] . \qquad (1.15)$$

The variables associated with the vertical links are integrated out in the limit in which the length $a_0$ of the vertical links approaches zero, so the hamiltonian depends only on the variables associated to horizontal links of the lattice. The Heisenberg equations of motion of this hamiltonian, and the constraints that are satisfied identically, provide a quantized and regularized version of all the classical Yang-Mills equations.

In addition to the hamiltonian, the quantum theory also requires specification of the configuration space. In sect. 5, the "horizon function" $G(A^{tr})$ is defined. It describes the fundamental modular region $\Lambda$ of QCD in the minimal Coulomb gauge on a *periodic* lattice in the infinite-volume limit by the condition $G(A^{tr}) \leq 0$ . (Its expression on a finite lattice is not known.) The original derivation of this function in the continuum [9] and lattice theory [10] has been improved recently [11], but it still cannot be regarded as rigorous. However, it has recently been discovered [12,13], following the work of Maggiore and Schaden [14], that $G(A^{tr})$ vanishes identically, $G(A^{tr}) = 0$ for all $A^{tr}$, on a finite lattice with *free boundary conditions,* as is explained in sect. 5 (although not of course on a finite periodic lattice). This is just what is wanted, because in the infinite-volume limit of the periodic lattice with the horizon condition, $\langle G \rangle / V = 0$, the distinction between free and periodic boundary conditions should be lost.

Thus the modified hamiltonian $H_{eff} = H_{coul} + \gamma_0 G$, differs from the canonical hamiltonian $H_{coul}$ by something that is formally zero.

The theory described by $H_{eff}$, eq. (1.14), is not a gauge theory unless the horizon condition (1.13) is satisfied, which fixes $\gamma$ to the critical value $\gamma_{cr} = \gamma(g_0)$. In sect. 6, it is shown that the horizon condition contradicts the usual perturbative expansion. We therefore introduce the new expansion in powers of $g_0$, eq. (1.11), and derive the self-consistency equation (6.22) which is an integral equation that determines the static color-Coulomb potential u.

In sect. 7, the new expansion (1.11) is everywhere substituted into $H_{eff}$, and a new zero-order lattice hamiltonian is obtained, eq. (7.20). To understand its physical features, we consider its continuum analog,

$$H_0 = H_{ho} + H_{ci} \tag{1.16}$$

$$H_0 = \Sigma_{k,\lambda}\, \omega_k\, a_\lambda^\dagger(k)\, a_\lambda(k) + (2V)^{-1} \Sigma_k\, \rho_{-k}^a\, v_k\, \rho_k^a, \tag{1.17}$$

where $\lambda$ is a two-valued polarization index, $\mu$ is a hadronic mass, $v_k \equiv k^2 u_k^2$, and $\rho_k^a$ and $u_k$ are the fourier transforms of the color charge density $\rho(x)^a$ and the static color-Coulomb potential defined in (1.3) and (1.11a) respectively. Here $H_{ho}$ is a harmonic oscillator hamiltonian with frequencies $\omega_k \equiv [k^2+\mu^4 u_k]^{1/2}$, and $H_{ci}$ is a static color-Coulomb interaction hamiltonian. The color charge density is given by $\rho = \rho_{dyn} + \rho_{ext}$, where $\rho_{dyn}$ is the color-charge density of gluons (and of dynamical quarks if any) and $\rho_{ext}$ may be the color-charge density of, say, a pair of external quarks.

The quantity $v_k$ or its fourier transform $v(x)$ has a clear intuitive meaning as a color-Coulomb interaction energy between color-charges. However $v(x)$ differs from the gauge-invariant energy $E(x)$ of a pair of external quarks at a separation x. The latter is the ground-state energy $H_0\Psi_0 = E(x)\Psi_0$, of the system with (approximate) hamiltonian $H_0$. It also appears in the rectangular Wilson loop of dimension T by $|x|$, which in the limit of large T has the value $W = \exp[-E(x)T]$. The two differ, $v(x) \neq E(x)$, because the external quarks distort the ground-state gluon wave-function $\Psi_0(A^{tr})$, by a QCD analog of vacuum polarization, and moreover $E(x) \leq v(x)$, as is shown in Appendix E.

Because the second term in (1.17) is quartic in the fields, the ground state wave function of $H_0$ cannot be calculated exactly. This prevents an exact determination of the kernel of the integral equation (6.22) that determines the static color-Coulomb potential $u(k)$. In sect. 8, an approximate kernel is used in this equation, with solution $w(k) = u^{(0)}(k)$. The approximate interaction energy $v^{(0)}(k) = k^2 w^2(k)$ may be written $v^{(0)}(k) = g_c{}^2(k)/k^2$, where $g_c(k)$ is a renormalized running coupling constant that characterizes the static color-Coulomb interaction (and the approximation made). It is not the gauge-invariant running coupling constant $g(k)$ of the renormalization-group, $g_c(k) \neq g(k)$, because $v(x) \neq E(x)$, as explained in the preceding paragraph. The asymptotic form at high momentum of $g_c(k)$ is calculated in sect. 8, and is compared in Appendix D with the renormalization group expression for $g(k)$. The two quantities have the same analytic form, but with different leading coefficients $b_c \neq b_{rg}$. Numerically $b_c$ gives the bulk of the contribution to $b_{rg}$, and it is verified that the coefficient $b_c$ agrees with contribution to $b_{rg}$ that comes from the conventional perturbative evaluation of $\langle M^{-1} \rangle$, as it should. At the end of Sect 8, the horizon condition is solved in the same approximation for the corresponding unrenormalized coupling constant $g_0(k)$.

In sect. 9, the asymptotic form at low momentum of the approximate static color-Coulomb potential $w(k) = u^{(0)}(k)$ is determined, and thus also of the approximate interaction energy $v^{(0)}(k) = k^2 w^2(k)$, which is found to have the asymptotic form at low momentum given by $v^{(0)}(k) \sim 1/k^{14/3}$. Such a highly singular infrared behavior implies confinement of color-charge in the approximate theory described by $H_0$ and $v = v^{(0)}$. For the low-momentum limit of the fourier transformed color-charge density is the total color charge $\rho^a(k = 0) = Q^a$, and consequently, by (1.17), there is an infrared contribution to the energy of any state given by,

$$(\Psi, H_0\Psi) \sim [2(2\pi)^3]^{-1} \int d^3k \; v(k) \; (\Psi, Q^2\Psi) \; . \qquad (1.18)$$

This diverges if v(k) is as singular as $1/k^3$, for all states $\Psi$ with non-zero color charge. The divergence cannot be canceled, because the remaining terms in (1.17) are positive. Thus the low momentum behavior found for $v^{(0)}(k)$ implies that all states with non-zero total color charge have infinite energy. This is a universal mechanism which confines all color charge in all representations whether carried by dynamical or external particles.

In sect. 10 we present our conclusions. We discuss to what extent the confining behavior just described can be expected to survive the approximations made, and the relation of the $1/k^{14/3}$ behavior of v(k) to a linearly rising potential $E(x) \sim K|x|$.

## 2. Restriction of the Configuration Space of a Hamiltonian System with a Large Number of Degrees of Freedom.

The Feynman-Kac formula expresses a quantum mechanical expectation value as the expectation value of a fictitious statistical mechanical ensemble. If H is the hamiltonian of the quantum mechanical system, and Z is the partition function of the statistical mechanical ensemble with transfer matrix T, then they are related by

$$\sum_i \exp( - \beta E_i ) = \mathrm{tr}[\exp(-\beta H)] = \lim_{M\rightarrow\infty} \mathrm{tr} T^M, \tag{2.1}$$

$$T = \exp(-\varepsilon H) \approx 1 - \varepsilon H, \quad \text{for } \varepsilon = M^{-1}\beta \tag{2.2}$$

$$Z \equiv \mathrm{tr} T^M. \tag{2.3}$$

This correspondence goes both ways: given H we may find T and Z, and given T or Z, we may find H.

In statistical mechanics there are well known equivalences between different ensembles such as the microcanonical and canonical. For systems with a large number of degrees of freedom, this translates back into an equivalence between different kinds quantum mechanical system. In some cases, this allows one to replace a system with a boundary by a much simpler system with a modified hamiltonian and no boundary.

Consider a system with N of degrees of freedom, labeled by an index k. In hamiltonian lattice gauge theory, the large-N limit is the infinite-volume limit. As a concrete example, the reader may suppose that the hamiltonian is quadratic

$$H = 2^{-1}\sum_{k=1}^{N} ( - \partial^2/\partial q_k^2 + E_k^2 q_k^2) \qquad (2.4)$$

where k is a discrete 3-momentum, and $E_k$ may be, for instance,

$$E_k^2 = |k|^2, \qquad (2.5a)$$

or

$$E_k^2 = \Lambda_k, \qquad (2.5b)$$

where $-\Lambda_k$ are the eigenvalues of the 3-dimensional lattice laplacian, or

$$E_k = 0. \qquad (2.5c)$$

However it will be seen that our results hold quite generally for a hamiltonian which is bounded below.

We shall be interested in the consequences of the restriction of gauge field configurations to a fundamental modular region $\Lambda$. As an example, the reader may suppose that the fundamental modular region is an ellipsoid defined by

$$G(q) \equiv \sum_{k=1}^{N} q_k^2/c_k^2 - N \leq 0, \qquad (2.6)$$

where the $c_k$ are all of order unity, but again our results will be seen to hold very generally. The wave function $\Psi(q)$ takes values inside this ellipsoid and is required to vanish on the boundary,

$$\Psi(q) = 0 \quad \text{for } G(q) = 0. \qquad (2.7)$$

For finite N it is difficult in general to find the eigenvalues of H with this ellipsoidal boundary condition, even for $E(k) = 0$.

This quantum mechanical problem corresponds to the partition function

$$Z_\Lambda = \int_\Lambda \prod_{k,t} dq_{k,t} \exp[ - S(q) ] \qquad (2.8)$$

$$S(q) = 2^{-1} \sum_{k,t} [ \varepsilon^{-1} (q_{k,t+1} - q_{k,t})^2 + \varepsilon E_k^2 q_{k,t}^2 ], \qquad (2.9)$$

where t = 1,... M labels "time" slice. The region of integration is limited by the M constraints

$$G(q_t) \equiv \sum_k q_{k,t}^2/c_k^2 - N \leq 0, \qquad (2.10)$$

$t = 1,\ldots M$, as indicated by the subscript $\Lambda$ on the integral (2.8).

To evaluate $Z_\Lambda$ we write

$$\int_\Lambda \prod_{k,t} dq_{k,t} = \int \prod_{k,t} dq_{k,t} \prod_t \theta[\, - G(q_t)\, ], \qquad (2.11)$$

and we represent the step-function $\theta$ by its fourier expansion

$$\theta[\, - G(q_t)] = (2\pi)^{-1} \int_{-\infty}^{\infty} dy_t\, (\varepsilon + iy_t)^{-1} \exp[\, - iy_t\, G(q_t)\, ]. \qquad (2.12)$$

The integrand is an analytic function of the complex variables $z_t = x_t + iy_t$. For each t, we deform the contour of integration into the right-hand half plane, $x_t > 0$, which gives

$$Z_\Lambda = \int \prod_t [(2\pi i)^{-1} dz_t] \exp[\, W(z) - \Sigma_t \ln z_t\, ], \qquad (2.13)$$

where $z \equiv (z_1,\ldots z_M)$. The path of integration of $z_t$ runs parallel to the imaginary axis in the right-hand half plane, and W(z) is defined by

$$\exp[\, W(z)\, ] \equiv \int \prod_{k,t} dq_{k,t}\ \exp[- S(q) - \Sigma_t\, z_t\, G(q_t)\, ]\,. \qquad (2.14)$$

We look for a multi-dimensional saddle point at a critical value, $z_t = \alpha_t$ which is real and positive for each t, $\alpha_t > 0$. The critical value satisfies

$$\partial W(\alpha)/\partial\alpha_t - \alpha_t^{-1} = - \langle G(q_t)\rangle_\alpha - \alpha_t^{-1} = 0, \qquad (2.15)$$

where $\alpha \equiv (\alpha_1,...\alpha_M)$, and the expectation value is calculated in the ensemble defined by (2.14) with $z_t = \alpha_t$. The $\alpha_t$ are all equal $\alpha_s = \alpha_t$. The inverse "width" of the multi-dimensional saddle is described by the matrix

$$\partial^2 W(\alpha)/\partial\alpha_s\, \partial\alpha_t + \delta_{s,t}\alpha_t^{-2}$$

$$= \langle G(q_s)\, G(q_t)\rangle_\alpha - \langle G(q_s)\rangle_\alpha \langle G(q_t)\rangle_\alpha + \delta_{s,t}\alpha_t^{-2} \qquad (2.16)$$

in which the covariance matrix of the statistical mechanical observables $G(q_t)$ appears. The observable $G(q_t)$ is a bulk quantity of the order of the number of degrees of freedom N, so $\langle G(q_t)\rangle_\alpha$ is of order N, and its covariance is of the order $\delta_{s,t}N$. Consequently the width of the saddle is of order $N^{-1/2}$ in each variable, and a saddle-point evaluation becomes exact in the limit of large N.

We now prove that (2.15) has exactly one solution with real positive $\alpha$, where $\alpha$ is the common value of the $\alpha_t$. As the $\alpha_t$ are all equal, we may replace these M equations by the one equation

$$F(\alpha) \equiv M^{-1}\langle B\rangle_\alpha + \alpha^{-1} = 0 \qquad (2.17)$$

where

$$B \equiv \textstyle\sum_t G(q_t) \qquad (2.18)$$

and M is the number of time slices. We have

$$F'(\alpha) = - \{ M^{-1} [\langle B^2\rangle_\alpha - \langle B\rangle_\alpha{}^2] + \alpha^{-2} \} < 0, \qquad (2.19)$$

so $F(\alpha)$ is monotonically decreasing. Moreover $G(q_t)$ and hence B is bounded below, so $F(0) = +\infty$ by (2.17). At $\alpha = +\infty$, the partition function is dominated by the lower bound of $H(q_t)$ which by (2.10) is -N, so $F(+\infty) = - NM < 0$. Thus there is precisely one positive solution $\alpha$.

We conclude that in the large-N limit, the saddle point approximation becomes exact. This gives

$$\lim_{N\to\infty} Z_\Lambda = \lim_{N\to\infty} Z_\alpha, \qquad (2.20)$$

where $Z_\alpha$ is defined by

$$Z_\alpha \equiv \int \prod_{k,t} dq_{k,t} \ \exp [- S(q) - \alpha \sum_t G(q_t) ], \qquad (2.21)$$

provided that $\alpha$ has the critical value determined by

$$F(\alpha) = \langle G(q_t)\rangle_\alpha + \alpha^{-1} = 0, \qquad (2.22a)$$

where the expectation value is calculated in the ensemble $Z_\alpha$. Moreover the last equation has exactly one positive real solution.

Remarks: 1. The critical value $\alpha$ is of order unity, whereas $G(q_t)$ is a bulk quantity of order N. Consequently in the large-N limit, $\alpha$ may be neglected in eq. (2.22a) which simplifies to

$$\lim_{N\to\infty} N^{-1} \langle G(q_t)\rangle_\alpha = 0. \qquad (2.22b)$$

A further property of a bulk quantity is that its fluctuations are negligible, of relative order $N^{-1/2}$, so the probability distribution of $N^{-1}G(q_t)$ gets concentrated at its mean value,

$$N^{-1}[G(q_t) - \langle G(q_t)\rangle_\alpha] \approx O(N^{-1/2}) \qquad (2.23a)$$

which we write

$$G(q_t) \approx \langle G(q_t)\rangle_\alpha. \qquad (2.23b)$$

Consequently when (2.22b) holds, the probability distribution of $G(q_t)$ gets concentrated at the boundary of the fundamental modular region, at $G(q_t) = 0$ in the large-N limit. This is slightly counter-intuitive because the probability distribution of the variables $q_{k,t}$ remains peaked at $q_{k,t} = 0$, and $G(0) = - N$. The paradox is easily resolved however, because it is the fluctuation of the $q_{k,t}$ (the mean value of $q_{k,t}{}^2$) that contributes to the mean value of $G(q_t)$. We call G the "horizon function", and eq. (2.22b) the "horizon condition", where "horizon" refers to the boundary of the fundamental modular region.

2. The result (2.20) is an exact analog of the relation between the micro-canonical and the canonical ensemble.

3. We emphasize that only very general properties of H and G have been used in the above saddle-point evaluation. The above results hold for a wide variety of systems with boundary conditions, and is not at all restricted to the quadratic hamiltonian which is presented as an illustrative example only.

In the proof above that the equation $F(\alpha) = 0$ possesses a real positive solution $\alpha_{cr} > 0$, it was implicitly assumed that $F(\alpha)$ is analytic. However $F(\alpha)$ may develop a non-analyticity in the large-N limit, and $\alpha_{cr}$ may approach zero, as illustrated in fig. 1. [Indeed, in replacing (2.22a) by (2.22b), we dropped the term $\alpha^{-1}$ which was used to show that $F(0) > 0$ in the proof that the equation $F(\alpha) = 0$ does in fact possess a solution for positive $\alpha$.] There are two possibilities:

case (a) $\langle G(q_t) \rangle_0 \leq 0.$ (2.24a)

(We write $\langle G(q_t) \rangle_0$ for $\langle G(q_t) \rangle_\alpha$ at $\alpha = 0_+$.) In the large-N limit, the fluctuations of $G(q_t)$ are negligible, so in the ensemble characterized by $\alpha = 0_+$, the probability distribution of $G(q_t)$ gets concentrated on the surface $G(q_t) = \langle G(q_t) \rangle_0 \leq 0$ that is contained within the fundamental modular region. In this case the restriction to the fundamental modular region is vacuous, and we have $\alpha_{cr} = 0$.

case (b) $\langle G(q_t) \rangle_0 > 0.$ (2.24b)

In this case, the equation $\langle G(q_t) \rangle_\alpha = 0$ holds for some $\alpha = \alpha_{cr} > 0$. The fluctuations of $G(q_t)$ are negligible in the large-N limit, so in the distribution characterized by $\alpha = \alpha_{cr} > 0$, the probability distribution of $G(q_t)$ gets entirely concentrated on the surface defined by $G(q_t) = 0$, namely on the boundary of the fundamental modular region.

These two cases correspond to different phases in the large-N limit, for which $\langle G(q_t) \rangle$ may serve as an order parameter. We call case (a), in which the restriction to the fundamental modular region is vacuous, the perturbative phase, and we call case (b), in which the probability distribution of $G(q_t)$ gets concentrated on the boundary of the fundamental modular region, the horizon phase. We shall find that in QCD, the horizon phase corresponds to the phase of confined gluons.

In the horizon phase, the entropy is stronger than the energy and pushes the system out so that the probability distribution of the horizon function is concentrated on the boundary of the fundamental modular region. On the other hand, in the perturbative phase, the energy is stronger so the probability that $G(q_t)$ attains its value on the boundary of the fundamental modular region is negligible.

The equivalence between the quantum mechanical and the statistical mechanical system, is obtained in the limit in which the parameter $\varepsilon = M^{-1}\beta$ approaches zero, where $\varepsilon$ is the parameter that defines the length of the "time" interval in (2.9). For small $\varepsilon$, the solution to the horizon condition gives a value of $\alpha$ that is of order $\varepsilon$. To see this, observe that the canonical ensemble $Z_\alpha$, defined by (2.21) with

$$\alpha = \varepsilon \, \gamma \qquad (2.25)$$

corresponds to the quantum mechanical system with effective hamiltonian

$$H_{eff} = H + \gamma G. \qquad (2.26)$$

In the large-N limit, the parameter $\gamma$ is determined by the "horizon condition" (2.22b) which (in the zero-temperature limit) corresponds to the condition

$$F(\gamma) \equiv (\Psi_0(\gamma), G(q)\, \Psi_0(\gamma)) = 0, \qquad (2.27)$$

where $\Psi_0$ is the ground state wave function of the hamiltonian $H_{eff}$. This is independent of $\varepsilon$, so (2.25) is the correct scaling of $\alpha$.

The quantum mechanical system defined by $H_{eff}$ with the horizon condition (2.27) contains no boundary, but it is nevertheless equivalent in the large-N limit to the original system with hamiltonian H and boundary condition (2.7). The new system is much more convenient for calculations, just as the canonical ensemble is more convenient than the microcanonical ensemble.

To verify that the quantum-mechanical horizon condition (2.27) has at most one solution, we observe that $F(\gamma)$ is monotonic. For by first order perturbation theory we have

$$\partial F(\gamma)/\partial\gamma = - 2\, (\Psi_0, G\, (H_{eff} - \lambda_0)^{-1}\, G\, \Psi_0) < 0, \qquad (2.28)$$

where $\lambda_0 = \lambda_0(\gamma)$ is the vacuum energy of $H_{eff}(\gamma)$, and the inverse is taken on the subspace orthogonal to $\Psi_0$.

From (2.26) and (2.27) and the fact that an eigenvalue is stationary with respect to infinitesimal variations of its eigenfunction, we have

$$F(\gamma) = \partial\lambda_0(\gamma)/\partial\gamma. \tag{2.29}$$

The horizon condition (2.27) is thus equivalent to the statement that the vacuum energy is stationary at the critical value of $\gamma$. In quantum field theory, $\Gamma(\gamma) = \lambda_0(\gamma)$ is known as the effective action or effective potential.

For the concrete example defined by (2.4) and (2.6), one has

$$H_{eff} = 2^{-1} \Sigma_k [ - \partial^2/\partial q_k^2 + \omega_k^2 q_k^2 ] - \gamma N, \tag{2.30}$$

where

$$\omega_k^2 \equiv E_k^2 + 2\gamma c_k^{-2}. \tag{2.31}$$

This is the hamiltonian of a system of decoupled harmonic oscillators with eigenvalues

$$\lambda = \Sigma_k (n_k + 2^{-1})\omega_k - \gamma N \tag{2.32}$$

where $n_k = 0, 1 \dots$ .

Equation (2.27) reads

$$F(\gamma) = \Sigma_k c_k^{-2} (\Psi_0, q_k^2 \Psi_0) - N = \Sigma_k (2 \omega_k c_k^2)^{-1} - N, \qquad (2.33)$$

so the horizon condition, $F(\gamma) = 0$, gives

$$\Sigma_k [ 2 c_k ( c_k^2 E_k^2 + 2\gamma )^{1/2} ]^{-1} - N = 0, \qquad (2.34)$$

This equation has a solution for positive $\gamma$ if and only if the condition

$$\Sigma_k ( 2 c_k^2 E_k )^{-1} > N. \qquad (2.35)$$

is satisfied. If it does, then the system is in the horizon phase with a positive value of $\gamma$ satisfying (2.34). If this condition is not satisfied, then $\langle G(q)\rangle_\gamma$ is negative at $\gamma = 0$, and the system is in the perturbative phase and the boundary condition may be ignored. In the large-N limit, (2.31) and (2.34) provide a complete solution to the original eigenvalue problem with an ellipsoidal boundary condition.

## 3. Definition and Elementary Properties of the Minimal Lattice Coulomb gauge.

Consider a finite Euclidean periodic lattice which is cubic in the space directions. It will be convenient for taking the continuous-time limit of a Euclidean lattice to label Euclidean sites by two indices x and t, where x is a 3-vector with integer components that labels a spatial site, and t is integer-valued <u>Euclidean</u> time. Spatial, or horizontal, links of the lattice will be labeled by (xy, t) and time-like, or vertical, links by (x, t). Lattice configurations U are described by horizontal and vertical link variables denoted respectively by $U_{xy,t}$ and $h_{x,t}$, and local gauge transformations g are described by the site variables $g_{x,t}$. These variables are elements of the SU(n) group, and the reversed link corresponds to the inverse group element. The gauge-transform of the configuration (U, h) by a local gauge transformation g, is written

$$(U^g)_{xy,t} = g_{x,t}^\dagger\, U_{xy,t}\, g_{y,t}. \tag{3.1a}$$

$$(h^g)_{x,t} = g_{x,t}^\dagger\, h_{x,t}\, g_{y,t+1}. \tag{3.1b}$$

In order to choose a gauge which makes the horizontal link variables as close to unity as possible, we require a measure of the deviation of the horizontal link variables from unity. For this purpose we take the quantity

$$I(U) \equiv \Sigma_t\, \Sigma_{(xy)}\, n^{-1}\, \mathrm{Re}\, \mathrm{tr}(1 - U_{xy,t})\ , \tag{3.2}$$

where the sum extends over all time-slices t, and horizontal links (xy) of the lattice. It is positive, $I(U) \geq 0$, and vanishes, $I(U) = 0$, if and only if $U_{xy,t} = 1$ on every horizontal link. The restriction of this quantity to the gauge orbit through an arbitrary configuration U is given by the (Morse) function

$$F_U(g) \equiv I(U^g) = \Sigma_t\, \Sigma_{(xy)}\, n^{-1}\, \mathrm{Re}\, \mathrm{tr}(1 - g_{x,t}^\dagger\, U_{xy,t}\, g_{y,t})\ , \tag{3.3}$$

regarded as a function on the local gauge group $G$, for fixed U. The gauge-choice which makes the horizontal link variables as close to unity as possible, is achieved by gauge-transforming with the gauge transformation that makes this function a minimum. After this transformation, the minimum on each gauge orbit occurs at $g_{x,t} = 1$. The set $\Lambda$ of all configurations U with this property constitutes the fundamental modular region,

$$\Lambda \equiv \{U: \ F_U(1) \leq F_U(g) \text{ for all } g\}. \tag{3.4a}$$

Degenerate absolute minima occur only on its boundary and are identified topologically. For a lattice of finite volume, they form a lower dimensional set. It is shown below that this gauge choice falls into the class of lattice Coulomb gauges, and we call it the <u>minimal lattice Coulomb gauge</u>.

Because the minimizing function (3.3) is independent of the vertical link variables $h_{x,t}$, and is a sum over uncoupled time-slices, the minimization independent on each time slice, and the fundamental modular region is the topological product

$$\Lambda = \otimes_t \Lambda_t \tag{3.4b}$$

where $\Lambda_t$ is the fundamental modular region for spatial configurations $U_t$ on the time-slice t,

$$\Lambda_t \equiv \{U_t: \ F_{U_t}(1) \leq F_{U_t}(g_t) \text{ for all } g_t\}. \tag{3.4c}$$

The minimal Landau gauge in D Euclidean dimensions is defined by the same condition that holds on each time slice in the minimal Coulomb gauge in D spatial dimensions. Consequently the results of [11] for the minimal Landau gauge hold here on each time slice.

We now discuss the character of the minimum on the time slice t, temporarily suppressing the Euclidean time variable t. At any local or absolute minimum, the first variation of the minimizing function is stationary, which gives the so-called gauge condition, and the second variation is non-negative. To obtain the explicit form of these quantities, we write the local gauge transformation $g_x = \exp(\omega_x)$, where $\omega_x = \omega_x{}^a t^a$ is an element of the Lie algebra of the SU(n) group. Here the $t^a$ form an anti-hermitian basis of the defining representation of this Lie algebra. They satisfy $[t^a, t^b] = f^{abc} t^c$, and are normalized to $\mathrm{tr}(t^a t^b) = - 2^{-1}\delta^{ab}$. We have, to second order in $\omega$,

$$F_U(g) = F_U(1) - (2n)^{-1}\Sigma_{(xy)}$$

$$\times [\ \mathrm{tr}\{\ [U_{xy} - U_{xy}{}^\dagger\ ]\ (\omega_y - \omega_x + 2^{-1}[\omega_x, \omega_y])\ \}$$

$$+\ 2^{-1}\ \mathrm{tr}\{\ [U_{xy} + U_{xy}{}^\dagger\ ]\ (\omega_y - \omega_x)^2\ \}\ ]$$

$$= F_U(1) + (2n)^{-1}\Sigma_{(xy)}$$

$$\times [A_{xy}{}^c\ (\omega_y{}^c - \omega_x{}^c + 2^{-1} f^{abc}\omega_x{}^a\omega_y{}^b)$$

$$+ 2^{-1} (\omega_y^a - \omega_x^a) G_{xy}^{ac} (\omega_y^c - \omega_x^c) ]. \quad (3.5)$$

Here we have introduced the real horizontal link variables

$$A_{xy}^c \equiv - \mathrm{tr}[ (U_{xy} - U_{xy}^\dagger) t^c ], \quad (3.6)$$

which approach the classical connection in the continuum limit, and

$$G_{xy}^{ab} \equiv - 2^{-1} \mathrm{tr}[ (t^a t^b + t^b t^a ) ( U_{xy} + U_{xy}^\dagger ) ] . \quad (3.7)$$

We shall have occasion to also consider lattices with free boundary conditions. For this reason it is useful to introduce a notation which is inherently geometrical and holds for an arbitrary lattice which may be periodic or which has a boundary. For any site variable $\varphi_x$, the link variables

$$(\nabla\varphi)_{xy} \equiv \varphi_y - \varphi_x \quad (3.8)$$

$$(a\varphi)_{xy} \equiv (1/2) (\varphi_x + \varphi_y), \quad (3.9)$$

are defined for all links (xy), including those for which x or y may lie on a face, edge or vertex. (The lattice operation $\nabla$, which maps site variables to link variables, is obviously a lattice analog of the corresponding continuum quantity, called d in Cartan's notation.) As convenient, we also label links by $(xy) = (x, i)$, for $y = x + e_i$, where $e_i$ points in the positive i-direction, so

$$A_i(x) = A_{x,y} \qquad (3.10)$$

$$\nabla_i \varphi(x) \equiv (\nabla\varphi)_{xy} \qquad a_i \varphi(x) \equiv (a\varphi)_{xy} \qquad (3.11)$$

(These expressions are undefined for sites x where $y = x + e_i$ is not a site of the lattice.) Given two link variables that are group elements $V_{xy}$ and $W_{xy}$, we may form the product link variable $P_{xy} = V_{xy}W_{xy}$, etc.

The natural inner product on link variables provides a definition of the operator $\nabla^*$ which is the dual to $\nabla$,

$$(\nabla^* A, \varphi) \equiv (A, \nabla\varphi) = \Sigma_{(xy)} A_{xy} (\nabla\varphi)_{xy} = \Sigma_x (\nabla^* A)_x \varphi_x, \qquad (3.12a)$$

where the sums extend over all links and sites of the lattice. Whereas $\nabla$ maps site variables $\varphi_x$ into link variables $(\nabla\varphi)_{xy}$, its dual $\nabla^*$ maps link variables $A_{xy}$ into site variables $(\nabla^* A)_x$. We shall also use the notation $\nabla\cdot$ for the lattice divergence, $\nabla\cdot = -\nabla^*$

$$(\nabla\cdot A)_x \equiv - (\nabla^* A)_x . \qquad (3.12b)$$

The lattice divergence is defined for all sites x, and represents the sum of all link variables leaving the site x. (It is *not* formed simply of differences of link variables when x is a boundary point of the lattice.) Similarly the site variable $(a\cdot A)_x$ is defined by

$$(a\cdot A, \varphi) \equiv (A, a\varphi) = \Sigma_{(xy)} A_{xy} (a\varphi)_{xy} = \Sigma_x (a\cdot A)_x \varphi_x. \qquad (3.13)$$

With the help of these definitions, we rewrite the expansion (3.5), which is correct to second order, in the form

$$F_U(g) = F_U(1) + (2n)^{-1} [ (A, \nabla\omega) + (\nabla\omega, D(U)\omega) ]. \quad (3.14)$$

Here $[D(U)\omega]_{x,i}$ is the link variable defined by

$$D_i^{ac}\omega^c(x) \equiv G_i^{ac}(x)\nabla_i\omega^c(x) + f^{abc}A_i^b(x)a_i\omega^c(x) . \quad (3.15)$$

It is the infinitesimal change in the link variable $A_i^a(x)$ under an infinitesimal gauge transformation $\omega_x$. By analogy with the continuum theory, we call the operator D(U) that maps site variables into link variables "the lattice gauge-covariant derivative". We also define the lattice Faddeev-Popov matrix M(U) by

$$(\omega, M(U)\varphi) = (\nabla\omega, D(U)\varphi). \quad (3.16)$$

At a minimum, the first variation of the minimizing function is stationary, $-(A, \nabla\omega) = (\nabla\cdot A, \omega) = 0$, for all $\omega$, so configurations in the fundamental modular region $\Lambda$ satisfy the lattice transversality or lattice Coulomb-gauge condition

$$[\nabla\cdot A(U)]_x = 0. \quad (3.17)$$

At a stationary point of the minimizing function, the matrix of second derivatives is symmetric, so when A is transverse, the Faddeev-Popov matrix (3.16) is symmetric,

$$M(U) \equiv \nabla^* D(U) = D^*(U)\nabla \qquad (\text{for } \nabla\cdot A = 0), \qquad (3.18a)$$

which we write alternatively as

$$M(U) \equiv - \nabla\cdot D(U) = - D(U)\cdot\nabla \qquad (\text{for } \nabla\cdot A = 0). \qquad (3.18b)$$

Here $D^*(U) = - D(U)\cdot$ is dual to D(U), and maps link variables onto site variables, and we call $D(U)\cdot$ "the lattice gauge-covariant divergence". The Faddeev-Popov matrix is non-negative at a relative or absolute minimum. The two conditions together define the Gribov region Ω,

$$\Omega \equiv \{U: \ \nabla\cdot A(U) = 0 \text{ and } M(U) \geq 0 \}. \qquad (3.19)$$

The set Λ of absolute minima is contained in the set of relative mimima,

$$\Lambda \subset \Omega. \qquad (3.20)$$

When properly restricted to Λ, the Faddeev-Popov formula holds on each time-slice

$$1 = \int dg \prod \delta_\Lambda[\nabla\cdot A(U^g)] \, \det M(U^g). \qquad (3.21)$$

where

$$\delta_\Lambda[\nabla\cdot A(U^g)] \equiv \delta[\nabla\cdot A(U^g)]\, \chi_\Lambda(U^g). \tag{3.22}$$

Here $\chi_\Lambda(U)$ is defined only for configurations U that are transverse, and is given by $\chi_\Lambda(U) = 1$ for U in $\Lambda$, and $\chi_\Lambda(U) = 0$ otherwise. By virtue of this identity, Wilson's partition function (1.15) may be written in the minimal Coulomb gauge

$$Z = \int \prod\ dU_{x,i,t}\, dh_{x,t}\, \delta_\Lambda[\nabla\cdot A(U_t)]\, \det M(U_t) \exp[- S_W(U, h)\ ]. \tag{3.23}$$

Here $U_t$ refers to the configuration on the horizontal links of the time-slice t, and $S_W(U, h)$ is Wilson's Euclidean lattice action.

(As discussed in [13], the last expression is symbolic to the extent that, for simplicity, we have ignored the fact that on each time slice the gauge condition leaves a global or x-independent gauge transformation arbitrary. Consequently, on a lattice of spatial volume V, only V-1 gauge conditions $\nabla\cdot A(U)_x = 0$ are linearly independent, and moreover M(U) has a trivial nullspace consisting of constant (x-independent) eigenvectors. The determinant detM(U) is defined on the orthogonal space.)

## 4. Lattice Coulomb Hamiltonian

The Euclidean action may be decomposed into parts that lie within a given time-slice or link adjacent time-slices,

$$S_W(U, h) = \sum_t S_{t+1,t} \qquad (4.1a)$$

$$S_{t+1,t} = \beta_e \sum_{x,i} S_{x,i}(U_t, U_{t+1}, h_t) + \beta_m \sum_{x,ij} S_{x,ij}(U_{t+1}) \,. \qquad (4.1b)$$

where the variables $U_t$, and $h_t$ live on horizontal and vertical links. The sums extend over all vertical and horizontal plaquettes respectively and $S_{x,i}$ and $S_{x,ij}$ are the corresponding one-plaquette actions. For different vertical and horizontal intervals, the coefficients have different values given by

$$\beta_e \equiv 2na/(g_0^2\, a_0) \qquad \beta_m \equiv 2na_0/(g_0^2\, a), \qquad (4.2)$$

which come from dividing each plaquette action by the square of the area of the plaquette, namely $(a_0 a)^2$ and $(a^2)^2$ for vertical and horizontal plaquettes respectively, and multiplying by the 4-volume element $a_0 a^3$. (For purposes of renormalization, one may prefer to use different coupling constants $g_t$ and $g_s$ on the vertical and horizontal plaquettes, to avoid renormalizing the speed of light [16]. However we shall ignore this point here.)

Only the variables $h_t$ appear in $S_{t+1,t}$, but the variables $h_{t+1}$ do not. This is the lattice analog of the of absence of $\partial A_0/\partial t$ in the Lagrangian of the continuum theory. In the continuum theory, the variable $A_0$ is not a true dynamical variable, and variation with respect to $A_0$ gives the constraint known as Gauss's law. In this section, the integration over the variables $h_{x,t}$ that live on vertical links will be done explicitly in the limit of small $a_0$, and will give the lattice form of Gauss's law.

To find the Hamiltonian, we express the partition function in terms of the transfer matrix T by

$$Z = \mathrm{tr}(T^M), \tag{4.3}$$

where T is related to the Hamiltonian operator $H_{coul}$ by

$$T = \exp( - a_0 H_{coul}) = 1 - a_0 H_{coul} + O(a_0^2). \tag{4.4}$$

We now choose the variables $A_{x,i,t}{}^c$ defined in eq. (3.6) as coordinates for $U_{x,i,t}$ on each coordinate patch of the SU(n) group manifold and on each time-slice t, so the gauge condition $\nabla \cdot A = 0$ is linear in the coordinates. The discrete variable that labels coordinate patches will be suppressed. In terms of these variables we have

$$\int \Pi_{x,i}\ dU_{x,i,t}\ \Pi_x \delta_\Lambda[(\nabla \cdot A)_{x,t}] =$$

$$\int \Pi_{x,i}\ dA_{x,i,t}\ |\psi(A_{x,i,t})|\ \Pi_x \delta_\Lambda[(\nabla \cdot A)_{x,t}]$$

$$= \int_\Lambda dA^{tr}\ \Pi_{x,i}\ |\psi(A_{x,i,t}{}^{tr})|\ , \tag{4.5}$$

where $|\psi(A_{x,i,t})|$ is the Haar measure on the SU(n) group in terms of the coordinates $A_{x,i,t}$ [see (A.31)]. As indicated, the integration over $dA^{tr} \equiv \Pi^\alpha dA^{tr,\alpha}$ is restricted to the fundamental modular region $\Lambda$, where, for each time-slice t, the $A^{tr,\alpha}$ are a set of independent coordinates that parametrize the transverse components of $A_{x,i,t}$. They will be chosen to be the fourier coefficients of $A_{x,i,t}$. (The $A^{tr}{}_{x,i,t}$ are not independent variables on each link.)

From the partition function (3.23) and (4.1), one reads off the transfer matrix T,

$$(T\Psi)(A^{tr}) = N \int_\Lambda dB^{tr} \prod_{x,i} |\psi(B_{x,i}{}^{tr})| \, \det M(A_t{}^{tr}) \prod_x dg_x$$

$$\times \exp[- S_{t+1,t} ]\, \Psi(B^{tr}), \qquad (4.6)$$

where N is a normalization constant,

$$S_{t+1,t} \equiv \beta_e \Sigma_{x,i} S_{x,i}(A^{tr}, B^{tr}, g) + \beta_m \Sigma_{x,ij} S_{x,ij}(A^{tr}). \qquad (4.7)$$

and we have written

$$A_{x,i}{}^{tr} \equiv A_{x,i,t+1}{}^{tr}; \qquad B_{x,i}{}^{tr} \equiv A_{x,i,t}{}^{tr}; \qquad g_x \equiv h_{x,t}\, . \qquad (4.8)$$

The horizontal one-plaquette action $S_{x,ij} = S_{x,ij}(A^{tr})$ depends only on the horizontal variables within a single time slice. By comparison with (4.4), one sees that the horizontal plaquettes contribute a potential energy term to the hamiltonian, which, with $\beta_m = 2na_0(g_0{}^2a)^{-1}$, is given by

$$V(A^{tr}) \equiv 2n(g_0{}^2a)^{-1} \Sigma_{x,ij} S_{x,ij}(A^{tr}). \qquad (4.9)$$

We write

$$H_{coul} = K + V(A^{tr}), \qquad (4.10a)$$

or

$$T = \exp[ - a_0 V(A^{tr}) ] \exp[ - a_0 K ] . \qquad (4.10b)$$

The remaining part of the transfer matrix

$$T' \equiv \exp(- a_0 K) = 1 - a_0 K + O(a_0^2) \qquad (4.11)$$

acts according to

$$(T'\Psi)(A^{tr}) = N \int_\Lambda dB^{tr} \prod_{x,i} |\psi(B_{x,i}{}^{tr})| \det M(B^{tr}) \prod_x dg_x$$

$$\times \exp[- \varepsilon^{-1} \sum_{x,i} S_{x,i} ] \Psi(B^{tr}) . \qquad (4.12),$$

Here the vertical one-plaquette action $S_{x,i}$ is given by

$$S_{x,i} \equiv 1 - n^{-1} \mathrm{Re\,tr} [U^\dagger(A_{x,i}{}^{tr}) g_x^\dagger U(B_{x,i}{}^{tr}) g_y ], \qquad (4.13)$$

where $y \equiv x + e_i$, and

$$\varepsilon \equiv g_0^2 a_0 (2na)^{-1} \qquad (4.14)$$

is a small quantity that vanishes with $a_0$.

The evaluation of K, which turns out to be the kinetic energy operator, is done in Appendix A. The most transparent way to express K and the Coulomb Hamiltonian $H_{coul}$ is as quadratic forms. Let $\Phi(A^{tr})$ and $\Psi(A^{tr})$ be wave functions, where we have restored the superscript tr for transverse. The inner product is given by

$$(\Phi, \Psi) = \int_\Lambda dA^{tr}\, \sigma(A^{tr})\, \Phi^*(A^{tr})\, \Psi(A^{tr}), \qquad (4.15)$$

where $\sigma(A^{tr}) = \det\psi(A^{tr}) \det M(A^{tr})$, and $M(A^{tr}) = D^*(A^{tr})\nabla$. The subscript $\Lambda$ indicates that the integration is restricted to the fundamental modular region. With kinetic energy operator (A.54) and (A.55), and potential energy (4.9), the quadratic form corresponding to the Coulomb hamiltonian is given by

$$(\Phi, H_{coul}\Psi) = \int_\Lambda dA^{tr}\, \sigma(A^{tr}) \{\ [ g_0^2(2a)^{-1}] \sum_{x,i} (\mathcal{E}_{x,i}{}^a\Phi)^*(\mathcal{E}_{x,i}{}^a\Psi)$$

$$+ 2n(g_0^2\, a)^{-1}\Phi^* [\ \sum_{x,ij} S_{x,ij}\ ]\, \Psi\ \}, \qquad (4.16)$$

where $\Phi = \Phi(A^{tr})$, $\Psi = \Psi(A^{tr})$, and $S_{x,ij} = S_{x,ij}(A^{tr})$. Here $\mathcal{E}_{x,i}{}^a$ is the lattice color-electric field operator defined by

$$\mathcal{E}_{x,i,a} \equiv J_a{}^b(A^{tr}{}_{x,i})\, E_{x,i,b} \qquad (4.17)$$

where $J_a{}^b(A_{x,i})$ is given in (A.56) or (A.57), and $E_{x,i,b}$ acts according to

$$E\Psi = i\ [\ \partial\Psi/\partial A^{tr}\ -\ \nabla\, (D^*\nabla)^{-1}\, D^*\partial\Psi/\partial A^{tr}\ ]. \qquad (4.18)$$

The electric field operators satisfy the lattice non-Abelian Gauss's law constraint,

$$D^*E = D^*E = 0. \qquad (4.19)$$

The electric field operator may be written

$$E = E^{tr} - \nabla\varphi, \qquad (4.20)$$

where

$$E^{tr} = i\, \partial/\partial A^{tr}, \qquad (4.21)$$

and φ is the non-Abelian lattice Coulomb potential operator.

$$\varphi = i\, (D^*\nabla)^{-1}\, D^*\partial/\partial A^{tr}. \qquad (4.22)$$

This completes the calculation of the lattice Coulomb hamiltonian $H_{coul}$.

Remarks: 1. In case color charge is carried by other fields such as the quark field, then expression (4.22) for the color-Coulomb operator is modified to

$$\varphi = (D^*\nabla)^{-1}\, (\, i\, D^*\partial/\partial A^{tr} + \rho_{qu}), \qquad (4.23)$$

where $\rho_{qu}$ is the quark color charge density operator. Gauss's law is modified to

$$- D^*E = D\cdot E = \rho_{qu}. \qquad (4.24)$$

Gluons contribute to the color-Coulomb operator by the charge density operator $iD^*\partial/\partial A^{tr} = -iD\cdot\partial/\partial A^{tr}$
2. We give a more explicit expression for the case of SU(2). In this case we have on a given link

$$U(A) = \eta\ [\ 1 - (A/2)^2\ ]^{1/2} + t^a A^a, \qquad (4.25)$$

where we have written $A^a$ for $A^{tr}{}_{x,i}{}^a$, and $\eta = \pm 1$ is a sign that labels coordinate patches on the SU(2) group manifold. From $U^\dagger dU = t^a \psi^a{}_c dA^c$, we obtain for the Maurer-Cartan form for the SU(2) group in these coordinates,

$$\psi(A)^a{}_c = \eta\ [\ 1 - (A/2)^2\ ]^{1/2}\ \delta^{a,c}$$

$$+\ 4^{-1}\ \eta\ [\ 1 - (A/2)^2\ ]^{-1/2}\ A^a A^c - 2^{-1} f^{a,b,c}\ A^b, \qquad (4.26)$$

where $f^{a,b,c}$ are the structure constants of the SU(2) group. From $\psi^a{}_c J_a{}^b = \delta_c{}^b$, we obtain

$$J_a{}^c = \eta\ [\ 1 - (A/2)^2\ ]^{1/2}\ \delta_a{}^c - 2^{-1} f^{a,b,c}\ A^b, \qquad (4.27)$$

$$(J^*J)^{b,c} = J_a{}^b\ J_a{}^c = \delta^{b,c} - 4^{-1} A^b\ A^c, \qquad (4.28)$$

and the Haar measure on a single link is

$$\det\psi = |\, 1 - (A/2)^2 \,|^{-1/2} \,. \qquad (4.29)$$

Equation (4.28) for J*J on a given link yields for the sum in (4.16),

$$\Sigma_{x,i}\, (E_{x,i}{}^a\Phi)^*(E_{x,i}{}^a\Psi) = \Sigma_{x,i}\, [\, \delta^{b,c} - 4^{-1} A^{tr}{}_{x,i}{}^b\, A^{tr}{}_{x,i}{}^c \,]$$

$$\times\, (E_{x,i}{}^b\Phi)^*(E_{x,i}{}^c\Psi), \qquad (4.30)$$

where $E_{x,i}{}^b$ is defined in (4.18).

3. We have chosen a lattice notation which agrees with the continuum notation for corresponding quantities. To obtain the continuum expression for the kinetic energy operator, it is sufficient to set J = 1, and correspondingly $\psi = J^{*-1}$. Our expression then agrees with the continuum hamiltonian of Christ and Lee [2]. In the continuum, the gauge-covariant derivative is given by $D_i{}^{a,c} = \delta^{a,c}\nabla_i + f^{abc}A^{tr}{}_i{}^b$, so the contribution of the gluons to the charge density operator which appears in the Coulomb potential is $i\, (D^*\partial/\partial A^{tr})^a = -\, i\, (D\cdot\partial/\partial A^{tr})^a = -\, i\, f^{abc}\, A^{tr}{}_i{}^b\, \partial/\partial A^{tr}{}_i{}^c$.

4. The kinetic energy operator K is not the Laplace-Beltrami operator of the metric G, which is the operator that would be obtained if $\sigma = \det\psi\, \det M$ were replaced by $\det^{1/2}G = \sigma\, \det^{-1/2}(D^*D)$, in our expression for K. Of course the measure of integration in (4.15) and (4.16) could be trivially changed from $\sigma$ into $\det G^{1/2}$, by rescaling the wave functions by the factor $\det^{1/4}(D^*D)$. But then the operator K would be changed into the Laplace-Beltrami operator of the metric G plus a non-local potential term involving $(D^*D)^{-1}$. The interesting point is that it is possible to write $H_{coul}$ without any non-locality associated with $\det(D^*D)$ or $(D^*D)^{-1}$, even though these quantities appear in intermediate stages of the calculation. The only non-locality $H_{coul}$ is associated with the inverse Faddeev-Popov operator $M^{-1}(A^{tr}) = D^*(A^{tr})\nabla$.

5. In the classical Yang-Mills action,

$$S = 2^{-1}\int d^4x\, F^{\mu\nu a}(\partial_\mu A_\nu{}^a - \partial_\nu A_\mu{}^a + g_0 f^{abc} A_\mu{}^b A_\nu{}^c) - 2^{-1} F^{\mu\nu a} F_{\mu\nu}{}^a),$$

variation with respect to $A_0$ produces Gauss's law, $D \cdot E = 0$. This is the classical analog of our lattice procedure of integrating out the variables that live on the vertical links, which produced the lattice version of Gauss's law.

6. We would have obtained the same lattice Coulomb hamiltonian if we had followed an alternative procedure which is the lattice analog of the continuum method of Christ and Lee [2]. These authors first calculated the hamiltonian $H_W$ in the "$A_0 = 0$" or Weyl gauge, where the continuum coordinates $A_i(x)$ are Cartesian, and where there is a residual invariance under time-independent gauge transformations. They then obtained the continuum Coulomb hamiltonian $H_{coul}$ by making the non-linear change of variables $A_i = g^{-1}A_i{}^{coul}g + g^{-1}\nabla_i g$, and then letting $H_W$ act on gauge-invariant wave-functions. (The use of Cartesian coordinates as an intermediate step was advantageous for these authors whose purpose was to quantize a classical theory. In the present article, Cartesian coordinates are not needed, because we are gauge fixing the theory which is already quantized and regularized by Wilson.) The lattice analog of the "$A_0 = 0$" gauge is the "$U_0 = 1$" gauge. The lattice hamiltonian in the "$U_0 = 1$" gauge is the Kogut-Susskind hamiltonian [15], which is of the form (4.16) and (4.17), with the following modifications: the variables $A_{x,i}{}^a$ independent on each link; $\sigma(A) = \det\psi(A)$; the color-electric field operator $E_{x,i,b}$ is given simply by

$$E_{x,i,b} = i\, \partial/\partial A_{x,i,}{}^{b}, \qquad (4.31)$$

instead of (4.18), and there is no restriction to a fundamental modular region. However in the "$U_0 = 1$" gauge, there is a residual gauge-invariance under time-independent gauge transformations, and wave-functions are required to satisfy Gauss's law as a subsidiary condition to satisfy gauge invariance. (The left-hand side of Gauss's law is the generator of time-independent gauge transformations.) This is an independent assumption in the "$U_0 = 1$" gauge, whereas in the Coulomb gauge, we have derived gauge-invariance in the form of Gauss's law, by integrating over the vertical links. The last sentence contains a slight over-simplification. The quantum mechanical partition function $Z = \mathrm{tr}\exp(-\beta H)$ is a trace, and this in fact corresponds to periodic boundary conditions in Euclidean time. Strictly speaking, the "$U_0 = 1$" gauge does not exist with periodic boundary conditions, because it assigns to unity the value of Wilson loops that close by periodicity. A correct procedure is to make all vertical links unity except those at a fixed Euclidean time, and one may show that integration over these last vertical links indeed effects the projection onto gauge-invariant states. However we have preferred not to first pass to the $U_0 = 1$ gauge in order to maintain the connection with "optimal gauge fixing", and because our statistical mechanical arguments do not allow a special role for a fixed Euclidean time slice.

7. For the fundamental modular region $\Lambda$ to be truly the quotient-space of configurations U modulo gauge transformations g, it is necessary to identify some points on the boundary $\partial\Lambda$ which are gauge-transforms of each other. This means that the wave-functions are required to assume the same value at these points. However, because the Coulomb gauge may be regarded as the "$U_0 = 1$" gauge, with the subsidiary condition that the wave-functions be gauge invariant, the wave functions that are eigenfunctions of the Coulomb hamiltonian satisfy this requirement automatically [5]. Thus we need not be concerned to identify these points explicitly, and it is sufficient to cut off the integration over $\Lambda$ at the boundary $\partial\Lambda$.

8. The Gauss's law constraint satisfied by $E_{x,i}{}^a$, the definition of $A^0$, eq. (1.7), and the Heisenberg equations of motion of $A = A^{tr}$ and $E^{tr}$ generated by the hamiltonian $H_{coul}$ provide a quantized and regularized form of all the classical Yang-Mills equations on a lattice of finite spatial extent.

## 5. Fundamental Modular Region and Effective Coulomb-gauge Hamiltonian

The Coulomb hamiltonian, (4.16), does not by itself give a complete expression of the dynamics of a gauge theory, because the space of wave functions must also be restricted to the fundamental modular region $\Lambda$. (Alternatively, a signed measure may be used, as discussed in [4] and [5]). An explicit expression for its boundary $\partial\Lambda$ is not known for a lattice of finite volume. Fortunately, a limiting expression for $\partial\Lambda$, valid at large volume, was derived for the minimal lattice Landau gauge on a periodic lattice in D Euclidean dimensions in sects. 2-4 of [11], which holds also for the minimal Coulomb gauge on a given time slice in D spatial dimensions.

(The horizon function which was designated H in [11] is designated G in the present article.)

Let $U = U(A^{tr})$ be a transverse configuration on a single time slice of a periodic lattice. Let $G(A^{tr})$ be defined by

$$aG \equiv \Sigma_{x,y} D_{x,i}^{ac} D_{y,i}^{ad} M^{-1}_{x,c;y,d}$$

$$- (n^2 -1) \Sigma_{(xy)} n^{-1} \text{Re tr } U_{xy} , \qquad (5.1)$$

where, a is the lattice spacing, and $M = D^*\nabla = \nabla^*D$ is the Faddeev-Popov matrix. Its inverse $M^{-1}$ is taken on the subspace orthogonal to the trivial null space consisting of constant (x-independent) vectors. The lattice difference operator $\nabla$ and the lattice gauge covariant derivative $D = D(A^{tr})$, and their duals $\nabla^*$ and $D^*(A^{tr})$ are defined in sect. 3. (The sum in the second term is the minimizing function (3.2), apart from an additive constant.) Let $\Lambda$ be the fundamental modular region, namely the set of configurations $U(A^{tr})$ which are absolute minima of the minimizing function I(U), eq. (3.2), on each gauge orbit, $I(U) \leq I(U^g)$. Let $\Lambda'$ be the subset of the Gribov region $\Omega$ defined by $G(A^{tr}) \leq 0$,

$$\Lambda' = \{A^{tr}: G(A^{tr}) \leq 0\} \cap \Omega . \qquad (5.2)$$

Let $Z_\Lambda$ and $Z_{\Lambda'}$ be the corresponding partition functions

$$Z_\Lambda \equiv \int_\Lambda \prod dA^{tr} \, \sigma \exp[ - S(U) ] \qquad (5.3)$$

and

$$Z_{\Lambda'} \equiv \int_\Omega \prod dA^{tr}\, \sigma\, \theta[\, - G(U)\, ]\, \exp[\, - S(U)\, ], \qquad (5.4)$$

where $\sigma = \det\psi \det M$, $\det\psi(A^{tr})$ is Haar measure, and S is a generic action. Then these partition functions (and the corresponding expectation values) are equal in the infinite-volume limit,

$$\lim_{V\to\infty} Z_\Lambda = \lim_{V\to\infty} Z_{\Lambda'}. \qquad (5.5)$$

Loosely we may say that $\Lambda = \Lambda'$ in the infinite-volume limit.

<u>Remarks</u>:

1. In the interior of the Gribov region $\Omega$, the horizon function $G(A^{tr})$ is finite, but approaches infinity as $A^{tr}$ approaches a generic point on the boundary $\partial\Omega$. <u>Proof</u>: By definition of $\Omega$, the Faddeev-Popov matrix $M(A^{tr})$ is strictly positive (apart from the trivial null space) for all $A^{tr}$ in $\Omega$, and it develops a (non-trivial) null eigenvalue on the boundary of $\Omega$.

2. For all $U(A^{tr})$ in $\Omega$, the horizon function G(U) is bounded from below by its value at the vacuum configuration $U_{xy} = 1$, namely

$$G(U) \geq G(1) = -\,(n^2-1)VD, \qquad (5.6)$$

where V is the spatial volume and D is the number of spatial dimensions. (The quantity $F \equiv (n^2-1)VD$ is the total number of degrees of freedom of link variables on the lattice.)

Proof: The first term in (5.1) is positive for all configurations U in the Gribov region, because the Faddeev-Popov operator is positive there. This term attains its minimum value, zero, for the vacuum configuration $U_{xy} = 1$, because $D_{x,i}(U)$ is the lattice difference operator at $U = 1$, $D_{x,i}(1) = \nabla_i$, where $\nabla_i$ is the lattice difference operator, and on a periodic lattice

$$\Sigma_x (\nabla_i f)_x = 0, \tag{5.7}$$

where $f_x$ is any site variable. The second term of G(U) is obviously also a minimum at $U_{xy} = 1$, where it has the stated value.

3. The horizon function $G(A^{tr})$ vanishes identically on a finite lattice with free boundary conditions for any transverse configuration $A^{tr}$. Proof: Let i and a be a pair of indices with the ranges i = 1,...D, and a = 1,...($n^2$-1), and let $f_{i,a}$ be the color-vector valued site variable defined by

$$f_{i,a,x}{}^b = x_i \, \delta_a{}^b. \tag{5.8}$$

This function is well defined on an infinite lattice or a finite lattice with free boundary conditions, but it is not well-defined on a finite periodic lattice. One has

$$[M(A^{tr})f]_{i,a} = - D(A^{tr})\cdot[\nabla f_{i,a}] = - D(A^{tr})\cdot(e_i \, e_a), \tag{5.9}$$

where $e_a$ is the color-vector with components $e_a{}^b = \delta_a{}^b$, and $e_i$ is the link variable with the value $\delta_i{}^j$ on links (x,j). With this expression for $[M(A^{tr})f]_{i,a}$ it is easy to verify that

$$G(A^{tr}) = \sum_{i,a} [\ (Mf_{i,a}, M^{-1} Mf_{i,a}) - (f_{i,a}, Mf_{i,a})\ ] = 0. \qquad (5.10)$$

This equation is false on a finite periodic lattice, as shown by (5.6), but it holds rigorously and identically on a finite lattice with free boundary conditions. To the extent that the boundary of the lattice becomes irrelevant in the infinite-volume limit, the equation $G(A^{tr}) = 0$ may be considered to hold in a formal sense on a periodic lattice of infinite volume when the horizon condition $V^{-1}\langle G \rangle = 0$ is imposed as a limiting condition at large volume.

4. It might be thought that the introduction of the term $\gamma_0 G$ into the hamiltonian would destroy renormalizability. Renormalizability is not well explored in the Coulomb gauge. However it should be noted that the analogous procedure in the Landau gauge does not destroy renormalizability for any value of $\gamma_0$. In that case, one adds $\gamma_0 G$ to the Euclidean action, and this non-local term is made local by an integral representation in terms of auxiliary ghost fields. It has been shown [14] for any value of $\gamma_0$, that in a perturbative expansion in the Landau gauge, the renormalization constants may be chosen independent of $\gamma_0$, and thus have the same value as in theory with $\gamma_0 = 0$. The basic reason is the preceding remark, which shows that G is, in essence, a boundary term.

The horizon function $G(A^{tr})$ is a bulk quantity of the order of the lattice volume, so the results of sect. 2 are valid in the infinite-volume limit. Consequently the statistical mechanical system described by the partition function $Z_{\Lambda'}$ is equivalent at large spatial volume to the quantum mechanical system described by the effective hamiltonian

$$H_{eff} = H_{coul} + \gamma_0 G, \tag{5.11}$$

where $H_{coul}$ is given in (4.16) and G in (5.1). In the formal sense of remark 3, we have $G \approx 0$ at infinite-volume, so in this formal sense nothing has been added to $H_{coul}$, and we have $H_{eff} \approx H_{coul}$.

To calculate with this hamiltonian one makes the usual rescaling

$$A^{tr} = g_0 A^{0,tr}, \tag{5.12}$$

Having taken into account the boundary of the fundamental modular region, we shall ignore other possible non-perturbative effects, such as the existence of different coordinate patches to cover the SU(n) group manifold, and the integration on $A^{0,tr}$ will be extended to infinity.

It is convenient to rescale the wave function according to

$$\Psi = \sigma^{-1/2} \Psi', \tag{5.13}$$

where $\sigma = \det(\psi M)$, which gives the inner product the simple form

$$(\Phi, \Psi) = \int \prod dA^{tr} \Phi'^{*} \Psi', \tag{5.14}$$

and the hamiltonian becomes

$$H_{eff}' = \sigma^{1/2} H_{eff} \sigma^{-1/2}. \tag{5.15}$$

The advantage of this transformation is that $\sigma = \det(\psi M)$ appears only in $H_{eff}'$ in the combination

$$\sigma^{-1}\, \partial\sigma/\partial A^{tr} = \partial[\mathrm{tr}\, \ln(\psi M)]/\partial A^{tr}$$

$$= \mathrm{tr}(M^{-1}\partial M/\partial A^{tr}) + \mathrm{tr}(\psi^{-1}\partial\psi/\partial A^{tr}), \qquad (5.16)$$

so it is only necessary to calculate the inverse $M^{-1}$ instead of the determinant, detM. The explicit form of the hamiltonian after this rescaling is given in Appendix B.

## 6. Color-Coulomb Potential

In the remaining sections A is rescaled according to $A = g_0 A^{(0)}$, and $\partial/\partial A = g_0^{-1}\partial/\partial A^{(0)}$, and the superscript on $A^{(0)}$ will be dropped. We also set the horizontal lattice spacing a to unity.

We define the 3-dimensional Faddeev-Popov propagator $C_x$ and its fourier transform $C(\theta)$ by

$$C_{x-y}\, \delta^{a,b} = \langle M^{-1}{}_{x,a;y,b}\rangle. \qquad (6.1)$$

$$C_{x-y} = V^{-1} \sum_\theta \exp[i\theta\cdot(x-y)]\, C(\theta), \qquad (6.2)$$

where M is the Faddeev-Popov matrix, and we use the same symbol for the fourier transform. In Appendix C we show that $C(\theta)$ is of the form

$$C^{-1}(\theta) = \Lambda(\theta) - q_i(\theta)\, \Sigma_{i,j}(\theta)\, q_j(\theta), \tag{6.3}$$

where

$$q_i(\theta) \equiv 2 \sin(2^{-1}\theta_i) \tag{6.4}$$

$$\Lambda(\theta) \equiv \sum_i q_{\theta,i}^2 \ . \tag{6.5}$$

The second term in (6.3) is a self-energy, so $\Sigma$ is a self-energy with momentum factors removed. We also show in Appendix C that horizon condition is equivalent to

$$[\, \Sigma(0)_{i,j} - \delta_{i,j} \,]^{TT} = 0. \tag{6.6}$$

We suppose that the integral for $\Sigma_{i,j}(0)$ is (infra-red) convergent, as will be verified a posteriori, (8.22) below. Cubic or rotational invariance gives

$$\Sigma_{i,j}(0) = \text{const.} \times \delta_{i,j}, \tag{6.7}$$

and the horizon condition (6.6) is equivalent to the statement

$$\Sigma_{i,j}(0) = \delta_{i,j}. \tag{6.8}$$

From (6.3) this gives the important result,

$$C^{-1}(\theta) = q_i(\theta) \, [\, \Sigma_{i,j}(0) - \Sigma_{i,j}(\theta) \,] \, q_j(\theta). \qquad (6.9)$$

Thus the restriction to the fundamental modular region expressed by the horizon condition implies that the Faddeev-Popov propagator, which transmits the static color-Coulomb potential, is highly singular at $k = 0$.

We now come to a crucial point. The last equation contradicts the usual perturbative expansion, according to which $\Sigma$ is of leading order $g_0^2$, whereas the left hand side is of leading order $g_0^0$. To find a different expansion in $g_0$ that is compatible with the last equation, we use the expression for the self-energy derived in Appendix C

$$\Sigma_{\theta,i,j} = g_0^2 \, [(n^2-1)V^3]^{-1} \sum_{\theta',\theta''} \cos(2^{-1}\theta'_i) \cos(2^{-1}\theta''_j)$$

$$\times f^{abc} f^{aed} \langle A_{\theta-\theta',i}^b A_{\theta''-\theta,j}^e (M^{-1})_{\theta',c;\theta'',d} \rangle + \ldots \, . \qquad (6.10)$$

where the dots represent higher order terms, and the fourier transform of $M^{-1}$ is defined by

$$(M^{-1})_{\theta',c;\theta'',d} = \sum_{x,y} (M^{-1})_{x,c;y,d} \exp[i(-\theta'\cdot x + \theta''\cdot y)], \qquad (6.11)$$

For $\theta' = \theta''$, this gives

$$V^{-1}(M^{-1})_{\theta',c;\theta',d} = \sum_z \exp(i\theta'\cdot z) \, V^{-1} \sum_x (M^{-1})_{x,c;x+z,d}. \qquad (6.12)$$

For each z, the quantity $V^{-1} \sum_x (M^{-1})_{x,c;x+z,d}$, being an average over all sites of the lattice, is of the form B/V, where B is a bulk quantity. Thus, as discussed at the end of sect. 2, if we separate out its vacuum-expectation or mean-field value, the remainder is of order $V^{-1/2}$.

$$V^{-1} \sum_x (M^{-1})_{x,c;x+z,d} = V^{-1} \langle \sum_x (M^{-1})_{x,c;x+z,d} \rangle + O(V^{-1/2})$$

$$V^{-1} \sum_x (M^{-1})_{x,c;x+z,d} = C_z \, \delta_{c,d} + O(V^{-1/2}) \qquad (6.13)$$

$$V^{-1} (M^{-1})_{\theta,c;\theta,d} = C_\theta \, \delta_{c,d} + O(V^{-1/2}). \qquad (6.14)$$

We conclude that in the infinite-volume limit the part of $M^{-1}$ that is diagonal in momentum space may be equated to its mean-value.

Consider the contribution of this diagonal part of $M^{-1}$ to the self-energy (6.10)

$$\Sigma_{\theta,i,j} = g_0^2 \, [(n^2-1)V^2]^{-1} \sum_{\theta',\theta''} \cos(2^{-1}\theta'_i) \cos(2^{-1}\theta''_j)$$

$$\times f^{abc} f^{aec} \langle A_{\theta-\theta',i}{}^b \, A_{\theta''-\theta,j}{}^e \rangle \, C_{\theta'} + \ldots \,. \qquad (6.15)$$

This is a disconnected part of $\Sigma_{\theta,i,j}$ in the sense that it is a product of vacuum expectation values. By translational invariance only the term with $\theta' = \theta''$ survives

$$\Sigma_{\theta,i,j} = n \, g_0^2 \, V^{-1} \sum_{\theta'} D_{\theta-\theta' i,j} \cos(2^{-1}\theta'_i) \cos(2^{-1}\theta'_j) \, C_{\theta'} + \ldots \,. \qquad (6.16)$$

where

$$D_{\theta,i,j}\,\delta^{b,e} \equiv \sum_y \exp(-i\theta\cdot y)\,\langle\, A_{x,i}{}^{b}\, A_{x+y,j}{}^{e}\,\rangle \tag{6.17}$$

is the equal-time gluon correlation function in momentum space. We substitute this expression into (6.9), and obtain

$$C_\theta{}^{-1} = n\, g_0{}^2\, V^{-1} \sum_{\theta'} q_i(\theta)\, [\, D_{\theta' i,j} - D_{\theta-\theta' i,j}\,]\, q_j(\theta)$$

$$\times \cos(2^{-1}\theta'_i)\, \cos(2^{-1}\theta'_j)\, C_{\theta'} + \ldots\, . \tag{6.18}$$

This is an integral equation for $C_\theta$. Because of the coefficient $g_0{}^2$, it blatantly contradicts the usual perturbative expansion, according to which C and D are of leading order $g_0{}^0$. To solve this equation for $C_\theta$, we require an expansion in powers of $g_0$ which is consistent with it. We therefore suppose that the Faddeev-Popov propagator $C_\theta$ is of leading order $g_0{}^{-1}$,

$$C_\theta = g_0{}^{-1}\, u_\theta + O(1). \tag{6.19}$$

Here $u_\theta$ is independent of $g_0$, and we call it "the static color-Coulomb potential". By (6.14), $M^{-1}$ acquires a mean-field contribution that is also of order $g_0{}^{-1}$,

$$V^{-1}(M^{-1})_{\theta,c;\theta',d} = g_0{}^{-1}\, u_\theta\, \delta_{\theta,\theta'}\, \delta_{c,d} + V^{-1}R_{\theta,c;\theta',d},$$

$$(M^{-1})_{x,c;y,d} = g_0^{-1} u_{x-y} \delta_{c,d} + R_{x,c;y,d}. \qquad (6.20)$$

We make the Ansatz that R and all other correlation functions are power series in $g_0$ with non-negative powers.

With this Ansatz, the terms in (6.18) represented by the dots are of higher order in $g_0$. Upon equating terms in (6.18) that are of leading order in $g_0$, we obtain an integral equation for u,

$$u^{-1}(\theta) = n\, (2\pi)^{-3} \int d^3\theta'\, q_i\, D^0_{i,j}(\theta')\, q_j\, \{ \cos(2^{-1}\theta'_i) \cos(2^{-1}\theta'_j)\, u(\theta')$$

$$- \cos[2^{-1}(\theta'+ \theta)_i] \cos[2^{-1}(\theta'+ \theta)_i]\, u(\theta'+ \theta) \}. \qquad (6.21)$$

where $q_i = q_i(\theta)$, and we have taken the infinite-volume limit. Here $D^0_{i,j}(\theta')$ is the equal-time gluon correlation function calculated to zero-order in $g_0$. This non-linear integral equation determines the static color-Coulomb potential $u(\theta)$, once $D^0_{i,j}(\theta)$ is given. It holds only in a non-abelian theory because the coefficient n which appears in front of the integral comes from the contraction of two structure constants, $f^{abc}f^{aec} = n\delta^{be}$.

Because of the difference that appears in the integrand, this equation remains finite in the critical limit, where it simplifies to

$$u^{-1}(q) = n\, (2\pi)^{-3} \int d^3k\, D^0(k)\, [\, q^2 - (q \cdot k)^2/k^2 \,]\, [\, u(k) - u(k+q) \,] . \qquad (6.22)$$

The horizon condition fixes $g_0 = g_0(\mu)$. The form (6.8), or equivalently,

$$\Sigma_{i,i}(0) = 3, \qquad (6.23)$$

is most appropriate because, by (6.3), it assures that $C_\theta$ is consistent with our expansion in every order. This gives, in leading order,

$$g_0^{-1} = n\,[3(2\pi)^3]^{-1} \int d^3\theta\, \Sigma_i\, D^0_{i,i}(\theta) \cos^2(2^{-1}\theta_i)\, u(\theta), \qquad (6.24)$$

where we have used (6.18) and (6.19) for $\Sigma_{i,i}(0)$.

## 7. Re-expanded Zero-order Hamiltonian

We require the kernel $D^0$ of the last equation which is given in (6.17). To evaluate this expectation-value to zero order in $g_0$ we need the ground-state wave-function $\Psi_0$ of the zero-order hamiltonian $H_0$. However to obtain the zero-order hamiltonian, we must use the new expansion, $(M^{-1})_{x,y} = g_0^{-1}u(x-y) + \ldots$ wherever $M^{-1}$ appears in the Hamiltonian $H_{eff}' = H_{coul}' + \gamma_0 G$. It appears in $H_{eff}'$ (B.1) or (B.3) in three places.

First, it appears in the Coulomb contribution to the color-electric field operator (4.18),

$$E = E^{tr} - \nabla\varphi, = i\, \partial/\partial A^{tr} - g_0\, \nabla M^{-1}\rho, \qquad (7.1)$$

where the color-electric charge density operator is defined by

$$\rho = i\, B^* \partial/\partial A^{tr}. \qquad (7.2)$$

Here we have replaced $D^*$ by $g_0 B^* = D^* - \nabla^*$, because $\nabla^* \partial/\partial A^{tr} = 0$. To leading order in $g_0$, the color-electric field and color-charge density operators are given by

$$E_{x,i}^a = i\, \partial/\partial A_{x,i}^{a,tr} - \nabla_{x,i} \Sigma_y\, u_{x-y}\, \rho_y^a + \ldots \qquad (7.3)$$

$$\rho_x^a = -\, f^{abc}\, 2^{-1} \Sigma_i\, [A_i^{tr,b}(x)\, E_i^{tr,c}(x) + A_i^{tr,b}(x-e_i)\, E_i^{tr,c}(x-e_i)\,] + \ldots \qquad (7.4)$$

where the dots represent terms that are higher order in $g_0$. In the new expansion, the Coulombic color-electric field operator has a zero-order piece, just like $E^{tr}$.

Second, $M^{-1}$ appears in the horizon function $\gamma_0 G$, eq. (5.1), which we write

$$\gamma_0 G = \gamma_0\, [\, g_0^2\, \Sigma_{x,y}\, B_{x,i}^{ab}\, B_{y,i}^{ac}\, M^{-1}{}_{x,b;y,c}$$

$$-\, (n^2-1)\, \Sigma_{x,i}\, n^{-1}\, \mathrm{Re\ tr}\, U_{x,i}\,]\,. \qquad (7.5)$$

Here we have replaced $D$ by $g_0 B$, because $\Sigma_x \nabla$ vanishes with periodic boundary conditions. We have

$$\gamma_0 G = \gamma_0 [ - 3 (n^2-1) V$$

$$+ g_0^2 \Sigma_{x,y} f^{adb} f^{aec} A_{x,i}^d A_{y,i}^e a_i^{(x)} a_j^{(y)} M^{-1}_{x,b;y,c} + ... ] \quad (7.6)$$

which gives

$$\gamma_0 G = \gamma_0 [ - 3 (n^2-1) V + n g_0 \Sigma_{x,y} A_{x,i}^b A_{y,i}^b a_i^{(x)} a_j^{(y)} u_{x-y} + ... ], \quad (7.7)$$

where $a_i^{(x)}$ is defined in (3.9). The order in $g_0$ of these terms depends on the order in $g_0$ assigned to $\gamma_0$. We pose

$$\gamma_0 = (2ng_0)^{-1} \mu^4, \quad (7.8)$$

where $\mu$ is of zero-order in $g_0$, so that $\gamma_0 G$ gives a non-trivial contribution to $H_0$. A perturbative expansion will be made in powers of $g_0$, at fixed $\mu$. Subsequently the horizon condition will be used to fix $g_0(\mu)$. This is similar to the substituting the solution of the renormalization-group equation for $g_0$ into the usual perturbative expansion.

By this assignment, $\gamma_0 G$ has the expansion,

$$\gamma_0 G = - 3 (n^2-1) (2 n g_0 )^{-1} \mu^4 V$$

$$+ 2^{-1}\mu^4 \Sigma_{x,y} A_{x,i}^b A_{y,i}^b a_i^{(x)} a_i^{(y)} u_{x-y} + ... \; . \quad (7.9)$$

The constant term of order $g_0^{-1}$ is irrelevant for the hamiltonian and will be dropped from $H_0$, although it contributes to the horizon condition $\langle G \rangle = 0$.

Finally $M^{-1}$ appears in the term

$$\mathrm{tr}(M^{-1}\partial M/\partial A_{x,i}{}^{a}) = \Sigma_{y,z}\,(M^{-1})_{y,b;z,c}\, f^{bdc}\, \partial(g_0 A^{d}{*}\nabla)_{z,y}/\partial A_{x,i}{}^{a}$$

$$+ \ldots \qquad (7.10)$$

in (5.16) that is present in $H_{eff}'$, and comes from absorbing the integration measure detM in the wave function. However because of the trace on the color indices, (7.10) has no contribution of order $g_0^{0}$. Thus to zeroth order we may replace $H_{eff}'$ by $H_{eff}$.

To write the hamiltonian explicitly, independent variables must be chosen so the transversality condition is satisfied identically. This is accomplished using the fourier components defined by

$$A^{tr}_{x,i} = (2V^{-1})^{1/2}\, \Sigma'_{\theta,\lambda}\, \varepsilon_{\theta,i}{}^{\lambda}\{ A_\theta{}^{c,\lambda} \cos[\theta\cdot(x + 2^{-1}e_i)]$$

$$+ A_\theta{}^{s,\lambda} \sin[\theta\cdot(x + 2^{-1}e_i)], \qquad (7.11)$$

$$E^{tr}_{x,i} = (2V^{-1})^{1/2}\, \Sigma'_{\theta,\lambda}\, \varepsilon_{\theta,i}{}^{\lambda}\{ E_\theta{}^{c,\lambda} \cos[\theta\cdot(x + 2^{-1}e_i)]$$

$$+ E_\theta{}^{s,\lambda} \sin[\theta\cdot(x + 2^{-1}e_i)], \qquad (7.12)$$

where the color index is temporarily suppressed, $\theta_i$ are the lattice momenta, $\theta_i = 2\pi n_i/N$, for $n_i = 1, 2,...N$, and the prime on the sum over $\theta_i$ indicates that the sum extends over half the Brillouin zone, and the polarization vectors $\varepsilon_{\theta,i}{}^{\lambda}$ will be specified shortly. To avoid a profusion of the superscript tr, the symbols A and E here and below refer only to the transverse components, which are the independent dynamical variables. These fourier components satisfy the canonical commutation relations

$$[E_\theta{}^{t,\lambda}, A_{\theta'}{}^{t',\lambda'}] = - [P_\theta{}^{t,\lambda}, A_{\theta'}{}^{t',\lambda'}] = i\, \delta_{\theta,\theta'}\, \delta^{t,t'}\, \delta^{\lambda,\lambda'} , \qquad (7.13)$$

where t = c or s.

We introduce the exponential fourier transforms

$$A^{b,tr}{}_{x,i} = V^{-1/2} \sum_{\theta,\lambda} \varepsilon_{\theta,i}{}^{\lambda} A_\theta{}^{b,\lambda} \exp[i\theta\cdot(x + 2^{-1}e_i)]$$

$$E^{b,tr}{}_{x,i} = V^{-1/2} \sum_{\theta,\lambda} \varepsilon_{\theta,i}{}^{\lambda} E_\theta{}^{b,\lambda} \exp[- i\theta\cdot(x + 2^{-1}e_i)], \qquad (7.14)$$

where the color index is restored, the sums now extend over the whole Brillouin zone, and

$$A_\theta{}^{\lambda} = 2^{-1/2} (A_\theta{}^{c,\lambda} - i\, A_\theta{}^{s,\lambda}); \quad A_{-\theta}{}^{\lambda} = (A_\theta{}^{\lambda})^* \qquad (7.15)$$

$$E_\theta{}^{\lambda} = 2^{-1/2} (E_\theta{}^{c,\lambda} + iE_\theta{}^{s,\lambda}); \quad E_{-\theta}{}^{\lambda} = (E_\theta{}^{\lambda})^*. \qquad (7.16)$$

These quantities satisfy the commutation relations

$$[\, E_\theta^{b,\lambda},\; A_{\theta'}^{b',\lambda'} \,] = i\, \delta_{\theta,\theta'}\, \delta_{b,b'}\, \delta_{\lambda,\lambda'}. \qquad (7.17)$$

The transversality conditions on $A^{tr}$, namely $\sum_i [\, A^{tr}_i(x) - A^{tr}_i(x-e_i) \,] = 0$, where $A^{tr}_i(x) \equiv A^{tr}_{x,i}$, and similarly for $E^{tr}$, are easily satisfied in momentum space because

$$\exp[i\theta\cdot(x + 2^{-1}e_j)] - \exp[i\theta\cdot(x - 2^{-1}e_j)] = iq(\theta_j)\exp(i\theta\cdot x) \qquad (7.18)$$

where $q(\theta_i) = 2\sin(2^{-1}\theta_i)$. It is sufficient to choose the polarization vectors $\varepsilon_{\theta,i}^{\lambda}$ to be a pair of orthonormal polarization vectors that are orthogonal to $q(\theta_i)$,

$$\sum_{i=1}^{3} q(\theta_i)\, \varepsilon_{\theta,i}^{\lambda} = 0 \qquad \lambda = 1, 2. \qquad (7.19)$$

To obtain the hamiltonian to zeroth order in $g_0$ we write expressions (4.16), (5.11), (7.3), (7.4) and (7.9) in terms of the canonical variables just defined, with the result

$$H_0 = H_{ho} + H_{ci}, \qquad (7.20)$$

where

$$H_{ho} \equiv 2^{-1} \sum_\theta \{E_{-\theta}^{b,\lambda}\; E_\theta^{b,\lambda}$$

$$+\; [\, \Lambda_\theta + \mu^4\, L_\theta^{\kappa,\lambda}\, u(\theta) \,]\, A_{-\theta}^{b,\kappa}\, A_\theta^{b,\lambda} \}\,, \qquad (7.21)$$

is a harmonic oscillator hamiltonian, and $H_{ci}$ contains the color-Coulombic interaction. The first term in (7.21) is the usual zero-order kinetic energy of the transverse color-electric field. The second term is a potential energy which is the sum of the usual zero-order potential energy of the color-magnetic field with coefficient $\Lambda_\theta = \Sigma_i q_{\theta,i}{}^2$ plus the zero-order potential energy of the horizon function $\gamma_0 G$ that contains the color-Coulomb potential $u(\theta)$. The matrix $L_\theta{}^{\kappa,\lambda}$ is defined by

$$L_\theta{}^{\kappa,\lambda} \equiv \Sigma_i \, \varepsilon_{\theta,i}{}^\kappa \cos^2(2^{-1}\theta_i) \, \varepsilon_{\theta,i}{}^\lambda \, ] \, . \qquad (7.22)$$

In the critical limit, $L_\theta{}^{\kappa,\lambda} = \delta^{\kappa,\lambda}$.

The color-Coulomb interaction hamiltonian $H_{ci}$ is the contribution of the zero-order Coulombic electric field (7.3),

$$H_{ci} \equiv 2^{-1} \, \Sigma_{x,i} \, (\nabla_i \varphi_x)^2 = \; 2^{-1} \, \Sigma_{x,i} \, (\nabla_i \, \Sigma_y \, u_{x-y} \, \rho_y{}^b)^2$$

$$H_{ci} = (2V)^{-1} \Sigma_\theta \, \rho_{-\theta}{}^b \, v_\theta \, \rho_\theta{}^b \, . \qquad (7.23)$$

where the color-electric charge $\rho_x{}^b$ is given in (7.4). It represents the color-Coulomb interaction energy with potential

$$v_\theta \equiv \Lambda_\theta \, u^2{}_\theta \, , \qquad (7.24)$$

The fourier components $\rho_\theta{}^b$ of the color-charge operator (7.4) are given by

$$\rho_x{}^a = V^{-1} \sum_x \exp(i\theta\cdot x)\, \rho_\theta{}^a \qquad (7.25)$$

$$\rho_\theta{}^a = -\, f^{abc} \sum_{\theta'} B_{\theta+\theta',\theta'}{}^{\lambda,\lambda'}\, A_{\theta+\theta'}{}^{b,\lambda}\, E_{\theta'}{}^{c,\lambda'}\,, \qquad (7.26)$$

where

$$B_{\theta,\theta'}{}^{\lambda,\lambda'} \equiv \sum_i \varepsilon_{\theta,i}{}^{\lambda} \cos[2^{-1}(\theta_i-\theta_i')]\, \varepsilon_{\theta',i}{}^{\lambda'}. \qquad (7.27)$$

In the critical limit this simplifies to $B_{\theta,\theta'}{}^{\lambda,\lambda'} = \varepsilon_\theta{}^{\lambda}\cdot\varepsilon_{\theta'}{}^{\lambda'}$. (In accordance with our convention, $A_\theta$ and $E_\theta$ are components of the *transverse* fields only.) The zero-momentum component of $\rho_\theta{}^a$ is the total color-charge operator

$$Q^a \equiv \rho_{\theta=0}{}^a = -\, f^{abc} \sum_\theta A_\theta{}^{b,\lambda}\, E_\theta{}^{c,\lambda} \qquad (7.28)$$

that satisfies the commutation relations of the Lie algebra of SU(n),

$$[Q^a, Q^b] = i\, f^{abc}\, Q^c. \qquad (7.29)$$

Upon diagonalizing the 2 by 2 matrix (7.22) one obtains harmonic oscillator frequencies $\omega_{\theta,\lambda}$, and eigenpolarizations $\varepsilon_{\theta,i}{}^{\lambda}$. We shall not do this explicitly, but note only that in the critical limit the harmonic oscillator frequencies are given by

$$\omega(k) = [\, k^2 + \mu^4\, u(k)\,]^{1/2}\,. \qquad (7.30)$$

If the gluons were free particles, this dispersion relation would contradict special relativity (unless u(k) is a constant), but if the gluons are confined by the color-Coulomb force, then there is no obvious contradiction.

With the introduction of the creation and annihilation operators,

$$a_\theta^{b,\lambda} \equiv (2\omega_{\theta,\lambda})^{-1/2} \left( \omega_{\theta,\lambda}\, A_\theta^{b,\lambda} - i\, E_{-\theta}^{b,\lambda} \right)$$

$$(a_\theta^{b,\lambda})^\dagger \equiv (2\omega_{\theta,\lambda})^{1/2} \left( \omega_{\theta,\lambda}\, A_{-\theta}^{b,\lambda} + i\, E_\theta^{b,\lambda} \right), \qquad (7.31)$$

which obey the commutation relations

$$[\, (a_\theta^{b,\lambda})^\dagger,\, a_{\theta'}^{b',\lambda'}] = \delta_{\theta,\theta'}\, \delta^{b,b'}\, \delta^{\lambda,\lambda'}, \qquad (7.32)$$

the harmonic oscillator hamiltonian takes the standard form

$$H_{ho} = \Sigma_{\theta,\lambda}\, \omega_{\theta,\lambda}\, [(a_\theta^{b,\lambda})^\dagger\, a_\theta^{b,\lambda} + 2^{-1}]. \qquad (7.33)$$

If quarks are present, the zero-order hamiltonian becomes

$$H_{0,tot} = H_0 + H_{0,qu}, \qquad (7.34)$$

where $H_{0,qu}$ is the free quark hamiltonian. Here $H_0$ is as above, but with color-charge density defined by

$$\rho = \rho_{gl} + \rho_{qu}, \qquad (7.35)$$

where $\rho_{gl}$ is given in (7.26) and $\rho_{qu}$ is the color-charge density of the quarks. The color-Coulomb interaction potential $v(\theta)$ couples at low momentum to the total color charge $Q^a$ of all colored particles. If the interaction potential $v_\theta$ is sufficiently strong at zero momentum then the energy of all states with non-zero color charge is infinite.

## 8. Determination of the Short Range Static Color-Coulomb Field

The static color-Coulomb potential, u(k), is the solution of the integral equation (6.22) whose kernel is the equal-time gluon correlation function $D^0$, eq. (6.17), evaluated in the ground state $\Psi_0$ of $H_0 = H_{ho} + H_{ci}$. However $\Psi_0$ cannot be calculated exactly because the Coulomb-interaction hamiltonian $H_{ci}$ is quartic in the field operators. In the remainder of this article we shall use the approximate* gluon correlator $D^{0(0)}$ evaluated in the ground state $\Psi_0{}^{(0)}$ of the harmonic oscillator hamiltonian $H_{ho}$. This introduces no error at high momentum because of asymptotic freedom, as is verified in Appendix D where the result obtained here is compared with the exact asymptotic expression for u(q) at large q calculated using the perturbative renormalization group. It is possible that at low momentum the only error is a renormalization of the mass scale.

In the present section we shall obtain the asymptotic form at high momentum of the solution

---

* We leave for another occasion to exhibit an improved ground-state wave-function obtained by a Bogoliubov transformation.

$$w(q) \equiv u^{(0)}(q) \qquad (8.1)$$

of the integral equation (6.22), in which the kernel $D^0$ is replaced by

$$D^{0(0)}{}_{i,j}(k) = [\ \delta_{i,j} - (k^2)^{-1} k_i k_j\ ]\ D^{0(0)}(k).$$

$$D^{0(0)}{}_k = (\ 2\omega_k)^{-1} = 2^{-1}\ [\ k^2 + \mu^4\ w(k)\ ]^{-1/2}\ , \qquad (8.2)$$

where we have used (7.30) for the frequencies $\omega_k$ of the harmonic oscillator hamiltonian $H_{ho}$. The last equation displays a reciprocal relation between the color-Coulomb potential $w(k) = u^{(0)}(k)$, and correlator of transverse gluons. The singularity (found in the following section) of w(k) at $k = 0$ suppresses $D^{0(0)}(k)$ at $k = 0$.

With this kernel, the integral equation (6.22) for w(q) reads

$$w^{-1}(q) = n\ [2(2\pi)^3]^{-1} \int d^3k\ [\ q^2 - (q \cdot k)^2/k^2\ ]$$

$$\times\ [\ k^2 + \mu^4\ w(k)\ ]^{-1/2}\ [\ w(k) - w(k+q)\ ]\ . \qquad (8.3)$$

Note that w(q) depends only on the magnitude of its argument. It is convenient to write $w(q^2)$ instead of w(q), and we have,

$$[\ q^2\ w(q^2)\ ]^{-1} = n\ [2(2\pi)^2]^{-1} \int_0^\infty dk\ k^2 \int_{-1}^{+1} dz\ (1 - z^2)$$

$$\times\ [\ k^2 + \mu^4\ w(k^2)\ ]^{-1/2}\ [\ w(k^2) - w(k^2+q^2+2kqz)\ ]. \qquad (8.4)$$

The Faddeev-Popov operator is positive inside the Gribov horizon, so we expect the solution $w(k^2)$ of this integral equation to be positive* for all k. We shall assume that there exists a unique positive solution.

We first find the color-Coulomb potential w(q) at momentum large compared to the mass scale, $q >> \mu$ (and also $q << a^{-1}$). As an Ansatz, suppose that at large q, the solution $w(q^2)$ differs from the free expression $w(q^2) = 1/q^2$ by a logarithmic power,

$$w(q^2) = C [ q^2 \ln^p(q^2/\mu^2) ]^{-1}. \qquad (8.5)$$

To determine the power p and the coefficient C, we define

$$f(q^2) \equiv [ q^2 w(q^2) ]^{-1}, \qquad (8.6)$$

and apply the operator $q\partial/\partial q$ to (8.4). This gives

$$q^2 f'(q^2) = - n [2(2\pi)^2]^{-1} \int_0^\infty dk\, k^2 \int_{-1}^{+1} dz\, (1 - z^2)$$

$$\times [ k^2 + \mu^4 w(k^2) ]^{-1/2} (q + kz)\, w'(k^2+q^2+2kqz) ]. \qquad (8.7)$$

We now write $k = qx$, and take $q >> \mu$, and we neglect the term $\mu^4 w(q^2x^2)$ compared to $k^2 = q^2x^2$. By (8.5) we have, at large q

---

* A numerical solution of this equation has been obtained recently which is indeed positive [17].

$$w'(q^2) = - C\ [ (q^2)^2 \ln^p(q^2/\mu^2) ]^{-1} [1 + p \ln^{-1}(q^2/\mu^2) ]$$

$$\approx - C\ [ (q^2)^2 \ln^p(q^2/\mu^2) ]^{-1}, \qquad (8.8)$$

and

$$w'(k^2+q^2+2kqz) \approx - C\{(q^2)^2 (x^2+1+2xz)^2 \ln^p[q^2(x^2+1^2+2xz)/\mu^2]\}^{-1}$$

$$\approx - C\{(q^2)^2 (x^2+1+2xz)^2 \ln^p(q^2/\mu^2)\}^{-1} \qquad (8.9)$$

This gives asymptotically at large q,

$$q^2 f'(q^2) = n\, C\, [2(2\pi)^2]^{-1} \int_0^\infty dx\, x \int_{-1}^{+1} dz\, (1 - z^2)\, (1 + xz)$$

$$\times \{ (x^2+1+2xz)^2 \ln^p[q^2/\mu^2]\}^{-1}. \qquad (8.10)$$

On the other hand, from (8.6) we have

$$q^2 f'(q^2) = C^{-1}\, p \ln^{p-1}(q^2/\mu^2). \qquad (8.11)$$

Equating the two expressions, we obtain p = 1/2, and

$$C^{-2} = n\, (2\pi)^{-2} \int_0^\infty dx\, x \int_{-1}^{+1} dz\, (1 - z^2)\, (1 + xz)\, (x^2+1+2xz)^{-2}$$

$$C^{-2} = n\, (2\pi)^{-2}\, (2/3). \qquad (8.12)$$

The color-Coulomb potential at high momentum is now completely determined,

$$w(q) = (3\pi^2/n)^{1/2} [\ q^2 \ln^{1/2}(q/\mu)\ ]^{-1}, \quad q >> \mu. \tag{8.13}$$

Its fourier transform is the color-Coulomb potential at short range in position space,

$$w(r) = (3\pi^2/n)^{1/2} (2\pi)^{-3} \int d^3q\ [\ q^2 \ln^{1/2}(q/\mu)\ ]^{-1} \exp(iq\cdot x)$$

$$w(r) = (3\pi^2/n)^{1/2} \{\ 4\pi\ r \ln^{1/2}[(\mu r)^{-1}]\ \}^{-1}, \quad r << \mu^{-1}. \tag{8.14}$$

The color-Coulombic electric field operator

$$E_c{}^a(x) = -\ \nabla \int d^3y\ w(x-y)\ \rho^a(y). \tag{8.15}$$

is expressed in terms of the gradient of this quantity,

$$-\nabla_i w(r) = (3\pi^2/n)^{1/2} [\ 4\pi\ r^2 \ln^{1/2}(\mu r)^{-1}\ ]^{-1} (x_i/r) \qquad r << \mu^{-1}. \tag{8.16}$$

The approximate color-Coulomb interaction energy,

$$v^{(0)}(q) \equiv q^2 w^2(q) = q^2\ [u^{(0)}(q)]^2, \tag{8.17}$$

is given at high momentum by

$$v^{(0)}(q) = (3\pi^2/n) [ q^2 \ln(q/\mu) ]^{-1}, \qquad q >> \mu_0, \qquad (8.18)$$

and at short range in position space,

$$v^{(0)}(r) = (3\pi^2/n) \{ 4\pi r \ln[ (\mu r)^{-1}] \}^{-1}. \qquad r << \mu^{-1} \qquad (8.19)$$

It is instructive to write this in the form

$$v^{(0)}(r) = g_c^2(r)/(4\pi r), \qquad (8.20)$$

where $g_c(r)$ is the Coulomb-gauge running coupling constant

$$g_c^{-2}(r) = b_c \ln[ (\mu r)^{-1}] \qquad r << \mu^{-1} \qquad (8.21a)$$

$$b_c = (3\pi^2)^{-1} n. \qquad (8.21b)$$

In Appendix D it is verified that this result agrees with the perturbative renormalization group.

We may use our solution $w(k)$ to evaluate $g_0(\mu)$, eq. (6.24). In the critical limit, it is given by

$$g_0^{-1} = n [3(2\pi)^3]^{-1} \int d^3k [k^2 + \mu^4 w(k)]^{-1/2} w(k). \qquad (8.22)$$

This integral converges in the infrared for the solution $w(k)$ at low momentum which will be obtained in the next section. However, in contrast to the integral equation for $w(k)$ which contains the difference [w(k) - w(q)] in the integrand, and is ultraviolet finite, the integral (8.22) has a logarithmic ultraviolet divergence. Instead of reverting to lattice kinematics, we calculate $g_0(\mu)$ to leading order in $a^{-1}$ by inserting a momentum cut-off in (8.22) at $\Lambda = a^{-1}$, where a is the lattice spacing. This gives

$$g_0(\mu)^{-1} = n\,(2\pi)^{-2}\,(2/3)\,C\,\ln^{1/2}(\Lambda^2/\mu^2), \qquad (8.23)$$

or, by (8.12)

$$g_0(\mu)^{-2} = b_c \ln(\Lambda/\mu), \qquad \Lambda \gg \mu, \qquad (8.24)$$

where $b_c$ is the coefficient (8.21) of the running coupling constant. Our expressions for unrenormalized coupling constant and for the renormalized coupling constant are consistent.

## 9. Determination of the Long-Range Color-Coulomb Potential

We next determine the solution w(q) of the integral equation (8.4) at low momentum. As an Ansatz, we suppose that for $q \ll \mu$, the color-Coulomb potential behaves like a power,

$$w(q) = B\,\mu^{s-2}\,q^{-s}. \qquad (9.1)$$

We change variables in (8.4) according to k = qx, and we now neglect $k^2$ compared to $\mu^4\,w(k^2)$. This gives

$$B^{-1} (q/\mu)^{s-2} = B^{1/2} (q/\mu)^{3-s/2}\ n\ [2(2\pi)^2]^{-1}\ I, \qquad (9.2)$$

where I is independent of q. We thus obtain $s = 10/3$. For this value of s, I is the numerical integral

$$I = \int_0^\infty dx\ x^{11/3} \int_{-1}^{+1} dz\ (1 - z^2)\ [\ x^{-10/3} - (x^2 + 1 + 2xz)^{-5/3}\ ]$$

$$I = 8\ \Gamma(8/3)\ \Gamma(2/3)\ \Gamma^{-1}(16/3). \qquad (9.3)$$

This gives

$$B^{-3/2} = n\ \pi^{-2}\ \Gamma(8/3)\ \Gamma(2/3)\ \Gamma^{-1}(16/3), \qquad (9.4)$$

$$u^{(0)}(q) = w(q) = B\ (\mu/q)^{4/3}\ q^{-2}, \qquad q << \mu. \qquad (9.5)$$

The approximate color-Coulomb potential w(q) is too singular at $q = 0$ to possess a fourier transform, but we may calculate the color-Coulomb electric field operator (8.15). If there is an external quark at the origin, for example, with color-electric charge $Q^a$, then according to (8.15), it produces a static color-electric field

$$E_i^a = -\ \nabla_i w(r)\ Q^a = (x_i/r)\ E_r\ Q^a \qquad (9.6)$$

where

$$-\ \nabla_i w(r) = -\ (2\pi)^{-3} \int d^3q\ iq_i\ \exp(iq\cdot x)\ B\ \mu^{4/3}\ q^{-10/3} \qquad (9.7)$$

is a finite integral, and the radial color-electric field $E_r$ at large r is given by

$$E_r = 3\,(16\pi^2)^{-1}\,3^{1/2}\,\Gamma(2/3)\,B\,\mu^{4/3}\,r^{-2/3} \qquad r >> \mu^{-1}, \qquad (9.8)$$

and falls off at large r like $r^{-2/3}$.

The color-Coulomb interaction energy at low q is given by $v^{(0)}(q) = q^2 w^2(q)$, or

$$v^{(0)}(q) = B^2\,\mu^{8/3}\,q^{-14/3}. \qquad q << \mu. \qquad (9.9)$$

The strong infrared singularity of this interaction energy provides the confining mechanism described in the introduction, eq. (1.18). It corresponds to a color interaction potential that grows at large r like $v^{(0)}(r) \sim r^{5/3}$.

## 10. Conclusion

We have developed a Hamiltonian scheme for the minimal lattice Coulomb gauge which includes the restriction of the configuration space to the fundamental modular region. This restriction implies that Faddeev-Popov propagator which transmits the color-Coulomb field is highly singular at k = 0. We derived a zero-order hamiltonian $H_0$, eqs. (1.16) or (7.20), that contains an instantaneous color-Coulomb interaction energy v(r) between color-charge, and which is the starting point for systematic corrections. An approximate evaluation of v(r) gives an expression $v^{(0)}(r)$ that agrees with the perturbative renormalization-group $[r \ln(\mu/r)]^{-1}$ at small r, and grows like $r^{5/3}$ at large r, thus exhibiting the qualitative features of a confining theory.

The question naturally arises to what extent these results are a gauge artifact. This term is somewhat of a misnomer since one does indeed have the privilege of fixing the gauge. The intended question is to what extent the results are an artifact of approximations made after fixing the gauge. In particular one may ask whether the rather unexpected $r^{5/3}$ behavior of the approximate color-Coulomb interaction energy $v^{(0)}(r)$ also characterizes the exact interaction energy v(r), and if so, and how this is to be reconciled with Seiler's upper bound on the energy E(r) of an external quark pair at separation r, $E(r) \leq Kr$, which holds at large r [18].

As a general remark it should be noted that the singular behavior in the infrared of the Faddeev-Popov propagator, exhibited by eq. (6.9), is an exact consequence of the restriction of the configuration space to the fundamental modular region. We believe that this is indeed the origin of a confining long-range color-Coulomb potential v(r) which appears in the zero-order hamiltonian $H_0$, and that this confining behavior survives the corrections due to the perturbation $H_{eff} - H_0$.

Whether or not the exact color-Coulomb interaction energy v(r) grows at large r like $r^{5/3}$ as does $v^{(0)}(r)$ requires further investigation to determine. The difference v(r) - E(r) is due to the distortion of the gluon wave function produced by the external quark pair, and is, in principle, a calculable effect, like vacuum polarization in QED. Moreover in Appendix E it is proven that in a world described by $H_0$, the inequality $E(r) \leq v(r)$ holds. The sign of this inequality and Seiler's asymptotic bound $E(r) \leq Kr$ on the energy $E(r)$ of an external quark pair at separation r [18], are obviously consistent with $E(r) \leq Kr < r^{5/3}$ at large r.

Finally we note that while a long-range instantaneous potential may explain confining behavior or an area law for a time-like Wilson loop, it says nothing directly about a space-like Wilson loop. In particular, the suppression at low momentum of the spatial components of the equal-time gluon correlator, (8.2), makes it hard to see in the Coulomb-gauge how confining behavior arises for a space-like Wilson loop. This is the same difficulty, alluded to in the Introduction, that occurs in the Landau gauge for all Wilson loops. That this difficulty is only apparent could be shown in the Coulomb gauge by a proof that Lorentz invariance is regained in the continuum limit.

The author recalls with pleasure stimulating and informative conversations with Gianfausto Dell'Antonio, Tony Duncan, T. D. Lee, Martin Schaden, Dieter Schuette, and Wolfhart Zimmermann. He is grateful to the organizers of the Fifth Meeting on Light-Cone Quantization and Non-perturbative QCD, Regensburg, Germany, June 20-23, 1995, for a stimulating conference that provided the inspiration to start this investigation. He is grateful to Professor Lee for sending him a preprint of ref. [5] before publication, and to Tony Duncan for bringing ref. [18] to his attention.

## Appendix A: Evaluation of the Kinetic Energy Operator

The evaluation of $T' = 1 - a_0 K$ to order $a_0$ will be done in a number of steps.

Step 1. Expansion about saddle-point.

We wish to evaluate the integral (4.12) for small values of $\varepsilon$. The variables $g_x = h_{x,t}$ that live on vertical links effect a gauge transformation on the link variables $U_{x,t}$ that lives on the time-slice t, for the one-plaquette action $S_{x,i}$ may be written

$$S_{x,i} = 1 - n^{-1} \mathrm{Re}\,\mathrm{tr}\,[U^{\dagger}(A_{x,i}^{tr})\, U^{g}(B_{x,i}^{tr})\,] . \qquad \text{(A.1)}$$

As $\varepsilon$ approaches zero, the integrand of (4.12) gets concentrated at the minimum of the action of the vertical plaquettes, which occurs at

$$U^{g}(B^{tr}) = U(A^{tr}), \qquad \text{(A.2)}$$

namely where $B^{tr}$ is a gauge-transform of $A^{tr}$. However a complete gauge-fixing has been done on each time-slice, so we conclude that at the minimum of the action

$$B^{tr} = A^{tr}, \qquad \text{(A.3)}$$

and

$$g = 1 \qquad \text{(A.4)}$$

hold. As $\varepsilon$ approaches zero, the integral (4.12) may be evaluated by Gaussian expansion around the minimum of the action, in which the dominant configurations are of order

$$B - A = O(\varepsilon^{1/2}) \qquad g = 1 + O(\varepsilon^{1/2}), \qquad \text{(A.5)}$$

as we shall see.

We set

$$B^{tr} = A^{tr} + \varepsilon^{1/2} v, \qquad \text{(A.6)}$$

and expand $\Psi(B^{tr})$ to order $\varepsilon$

$$\Psi(B^{tr}) = \Psi(A^{tr}) + \varepsilon^{1/2}\, v^{\alpha}\, \partial\Psi(A^{tr})/\partial A^{tr\alpha}$$

$$+ 2^{-1}\varepsilon\, v^{\alpha} v^{\beta}\, \partial^2\Psi(A^{tr})/\partial A^{tr\alpha}\partial A^{tr\beta}, \qquad \text{(A.7)}$$

where we have written v instead of $v^{tr}$ for simplicity. This gives

$$(T'\Psi)(A^{tr}) = I_0\, \Psi(A^{tr}) + I_1^{\alpha}\, \partial\Psi(A^{tr})/\partial A^{tr\alpha}$$

$$+ I_2^{\alpha\beta}\, \partial^2\Psi(A^{tr})/\partial A^{tr\alpha}\partial A^{tr\beta}, \qquad (A.8)$$

where the $I_a(A^{tr})$ are coefficients which we now evaluate.

Step 2. Evaluation of $I_0$.

Because $I_0\Psi$ is the only term in this equation that does not contain a derivative of $\Psi$, consistency with (4.11) requires that, with proper choice of normalization constant N, the coefficient $I_0$ be of the form

$$I_0(A^{tr}) = [\, 1 - a_0\, V'(A^{tr})], \qquad (A.9)$$

where $V'(A^{tr})$ is a (possible) potential energy contribution to K. By (4.12) we have

$$I_0 = N \int_\Lambda dB^{tr} \prod_{x,i} |\psi(B_{x,i}{}^{tr})|\, \det M(B^{tr}) \prod_x dg_x$$

$$\times \exp[- \varepsilon^{-1} \Sigma_{x,i}\, S_{x,i}\, ]\, , \qquad (A.10).$$

where $S_{x,i}$ is given in (A.1).

The easiest way to evaluate (A.10) is to undo the gauge-fixing. By (4.5), we have

$$I_0 = N \int \prod_{x,i}\, dB_{x,i}\, |\psi(B_{x,i})| \prod_x \delta_\Lambda[(\nabla\cdot B)_x]$$

$$\times \det M(B) \prod_x dg_x \exp[- \varepsilon^{-1} \Sigma_{x,i}\, S_{x,i}\, ]\, , \quad (A.11).$$

where

$$S_{x,i} = 1 - n^{-1}\mathrm{Re}\,\mathrm{tr}\,[U^{\dagger}(A_{x,i}^{tr})\, U^{g}(B_{x,i})] \,. \qquad \text{(A.12)}$$

We now use the invariance of the Haar measure $dB_{x,i}|\psi(B_{x,i})|$ under gauge transformation to write

$$I_0 = N \int \prod_{x,i} dB_{x,i}\, |\psi(B_{x,i})| \exp[- \varepsilon^{-1} \sum_{x,i} S_{x,i}]$$

$$\times \prod_x dg_x \prod_x \delta_\Lambda[(\nabla\cdot B^g)_x] \det M(B^g), \qquad \text{(A.13)}$$

where

$$S_{x,i} = 1 - n^{-1}\mathrm{Re}\,\mathrm{tr}\,[U^{\dagger}(A_{x,i}^{tr})\, U(B_{x,i})] \,. \qquad \text{(A.14)}$$

The integral over $dg_x$ gives unity by the Faddeev-Popov identity. From invariance of the Haar measure under left multiplication we obtain

$$I_0 = N \int \prod_{x,i} dB_{x,i}\, |\psi(B_{x,i})| \exp[- \varepsilon^{-1} \sum_{x,i} S_{x,i}] \qquad \text{(A.15)}$$

with

$$S_{x,i} = 1 - n^{-1}\mathrm{Re}\,\mathrm{tr}\, U(B_{x,i}) \,.$$

The last integral is independent of $A^{tr}$ and gives

$$I_0 = 1, \tag{A.16}$$

which determines the value of the normalization constant N,

$$N^{-1} = \prod_{x,i} \int dB_{x,i} \, |\psi(B_{x,i})| \exp\{- \varepsilon^{-1} \sum_{x,i} [ 1 - n^{-1}\mathrm{Re\, tr}\, U(B_{x,i}) ] \}. \tag{A.17}$$

and

$$V'(A^{tr}) = 0. \tag{A.18}$$

We conclude that the operator K contains no potential energy term.

Step 3. Evaluation of vertical plaquette to order $\varepsilon^{3/2}$.

There remains to evaluate the coefficients $I_1$ and $I_2$ defined in (A.8). We make the linear change of variable of integration from $B^{tr}$ to $v^{tr}$, defined by

$$B^{tr} \equiv A^{tr} + \varepsilon^{1/2} v^{tr}, \tag{A.19}$$

and we temporarily write A, B and v instead of $A^{tr}$, $B^{tr}$ and $v^{tr}$. We also set

$$g_x = \exp(\varepsilon^{1/2}\omega_x) \tag{A.20}$$

where $\omega_x = t^a \omega_x^a$ is an (anti-hermitian) element of the Lie algebra. From (4.12) we have

$$(I_1^{\alpha}, I_2^{\alpha,\beta}) = N \int dv \prod_{x,i} |\psi(B_{x,i})| \det M(B) \prod_x dg_x$$

$$\times \exp[- \varepsilon^{-1} \sum_{x,i} S_{x,i}] (\varepsilon^{1/2} v^{\alpha}, 2^{-1} \varepsilon v^{\alpha} v^{\beta}) . \quad \text{(A.21)},$$

where B stands for the right hand side of (A.19), and the one-plaquette action $S_{x,i}$ is given by (4.13). The limits of integration on v due to the restriction to the fundamental modular region are of order $\varepsilon^{-1/2}$. However to obtain a Gaussian expansion in powers of $\varepsilon^{1/2}$ about the saddle-point at v = 0, the limits of the v integration are made infinite.

Recall that we only need $(T'\Psi)(A^{tr})$ to order $\varepsilon$. Because the last factor of (A.21) contains explicit coefficients $\varepsilon^{1/2}$ and $\varepsilon$, it suffices to evaluate the first factor to order $\varepsilon^{1/2}$. To the order of accuracy required we have,

$$dg_x = d\omega_x. \quad \text{(A.22)}$$

(We ignore the overall normalization now because it is easier to determine it later.)

In order to evaluate to order $\varepsilon^{1/2}$ the first factor of (A.21), which contains the expression $\varepsilon^{-1} S_{x,i}$, we must evaluate the one-plaquette action $S_{x,i}$ to order $\varepsilon^{3/2}$. We write

$$S_{x,i} = 1 - n^{-1} \mathrm{Re\ tr} \exp F_{x,i} , \qquad (A.23)$$

where $F_{x,i}$ is an element of the Lie algebra which, according to (4.13), is given by

$$\exp F_{x,i} \equiv U^\dagger(A_{x,i})\, g_x^\dagger\, U(B_{x,i})\, g_y , \qquad (A.24)$$

where $y = x+e_i$. The leading term in $F_{x,i}$ is of order $\varepsilon^{1/2}$, and the leading term in $S_{x,i}$ is of order $F_{x,i}^2$, so to evaluate $S_{x,i}$ to order $\varepsilon^{3/2}$, it is sufficient to evaluate $F_{x,i}$ to order $\varepsilon$. For this purpose we write

$$\exp F_{x,i} = \exp F'_{x,i} \exp(\varepsilon^{1/2}\omega_y) , \qquad (A.25)$$

where

$$\exp F'_{x,i} \equiv U^\dagger(A_{x,i}) \exp( - \varepsilon^{1/2}\omega_x)\, U(B_{x,i}). \qquad (A.26)$$

We temporarily write A, B and F' for $A_{x,i}$, $B_{x,i}$ and $F'_{x,i}$. Upon taking hermitian conjugates we obtain

$$F'(B, A, -\omega_x) = - F'(A, B, \omega_x). \qquad (A.27)$$

It is convenient to change argument from A and B to

$$V_{x,i} \equiv 2^{-1} (A_{x,i} + B_{x,i}), \qquad (A.28)$$

and the variable $v_{x,i}$ previously introduced, $\varepsilon^{1/2}v_{x,i} = (B_{x,i} - A_{x,i})$. We have

$$\exp F' \equiv U^\dagger(V - 2^{-1}\varepsilon^{1/2}v) \exp( - \varepsilon^{1/2}\omega_x) U(V + 2^{-1}\varepsilon^{1/2}v). \qquad (A.29)$$

and

$$F'(V, - v, - \omega_x) = - F'(V, v, \omega_x), \qquad (A.30)$$

so $F(V, v, \omega_x)$ contains only odd powers of v and $\omega_x$. These variables appear with coefficient $\varepsilon^{1/2}$, so to obtain $F'(V, v, \omega_x)$ to order $\varepsilon$, it is sufficient to expand it up to terms linear in v and $\omega_x$, at fixed V.

For this purpose we write, for an element U(A) of the SU(n) group, considered as a function of coordinates $A^b$,

$$U^\dagger dU = \psi_c(A)\, dA^c = t^b\, \psi^b{}_c(A)\, dA^c, \qquad (A.31)$$

where b and c are color indices, the $t^b$ form an anti-hermitian basis of the Lie algebra of SU(n), and $\psi_c(A)dA^c$ is the (Lie-algebra valued) Maurer-Cartan one-form of the SU(n) group. The matrix $\psi^a{}_b(A)$ maps elements $dA^b$ of the cotangent space of link variables U(A) at A into Lie-algebra valued link variables. The Haar measure on the SU(n) group is given by $|\psi(A)| \equiv \det\psi(A)$. [Explicit expressions for these quantities in the case of the SU(2) group are given in eqs. (4.25) to (4.29).] To order $\varepsilon^{1/2}$ we have

$$U(V \pm 2^{-1}\varepsilon^{1/2}v) = U(V) [ 1 \pm 2^{-1}\varepsilon^{1/2}\psi_d(V)v^d ], \qquad (A.32)$$

so to order ε we have

$$\exp F' = [\, 1 + 2^{-1}\varepsilon^{1/2}\, \psi_d(V)\, v^d \,]\, U^\dagger(V)\, \exp(\, - \varepsilon^{1/2}\omega_x)\, U(V)$$

$$\times [\, 1 + 2^{-1}\varepsilon^{1/2}\, \psi_d(V)\, v^d \,]$$

$$= \; 1 + \varepsilon^{1/2}\, [\, \psi_d(V)v^c - U^\dagger(V)\omega_x U(V)\, ],$$

which gives, to order ε

$$F' = \varepsilon^{1/2}\, [\, \psi_d(V)v^d - U^\dagger(V)\omega_x U(V)\, ]\, . \qquad \text{(A.33)}$$

From (A.25) we obtain, to order ε,

$$\exp F = \exp\{\varepsilon^{1/2}\, [\, \psi_d(V)v^d - U^\dagger(V)\omega_x U(V)\, ]\, \}\, \exp(\varepsilon^{1/2}\omega_y)\, . \text{(A.34)}$$

By (A.23) we have, for the one-plaquette action, correct to order $\varepsilon^{3/2}$,

$$S_{x,i} = \varepsilon\, (4n)^{-1} \sum_b \{\, \psi^b{}_d(V_{x,i})v_{x,i}{}^c + \omega_y{}^b - [U^\dagger(V_{x,i})\omega_x U(V_{x,i})]^b \,\}^2 \,, \qquad \text{(A.35)}$$

where $y = x+e_i$, and we have used $\mathrm{tr}(t^a t^b) = -2^{-1}\delta_{a,b}$.

Let $D$(A) be the matrix that maps Lie-algebra valued site variables ω into the Lie-algebra valued link variable according to

$$t^c\,[\mathcal{D}(A)\omega]_{x,i}{}^c \equiv \omega_y - U^\dagger(A_{x,i})\omega_x U(A_{x,i}), \qquad (A.36)$$

where $y = x+e_i$. The right hand side is the gauge covariant difference between $\omega_y$ and $\omega_x$, where $\omega_x$ is transported to y by $U(V_{x,i})$. The operator $\mathcal{D}(A)$ is related to the "lattice gauge-covariant derivative D(A), introduced in (3.15) by the change of basis

$$[\mathcal{D}\,(A)\omega]_{x,i}{}^b \equiv \psi^b{}_a(A_{x,i})\,[D(A)\omega]_{x,i}{}^a, \qquad (A.37)$$

where $\psi^b{}_a(A_{x,i})$ is the Maurer-Cartan form of the SU(n) group defined in (A.31). The adjoint matrices $\mathcal{D}^*(A)$ and $\psi^*(A)$ are defined be means of the natural inner product of Lie-algebra valued link or site variables, as in (3.12). In particular $\mathcal{D}^*(A)$ maps Lie-algebra valued link variables into Lie-algebra valued site variables. To prove (A.37), we write $g_{x,y} \equiv \exp(\varepsilon^{1/2}\omega_{x,y})$, and we have, to order $\varepsilon^{1/2}$

$$\varepsilon^{1/2}\,[\ \omega_y - U^\dagger(A_{x,i})\omega_x U(A_{x,i})\ ]$$

$$= U^\dagger(A_{x,i})\,[\ g_x{}^\dagger U(A_{x,i})g_y - U(A_{x,i})\ ]$$

$$= U^\dagger(A_{x,i})\,[\ U(A^g{}_{x,i}) - U(A_{x,i})\ ]$$

$$= U^\dagger(A_{x,i})\,\{U[\ (A + \varepsilon^{1/2}D(A)\omega)_{x,i}\ ] - U(A_{x,i})\ \},$$

$$= \varepsilon^{1/2}\, t^b\, \psi^b{}_a(A_{x,i})\,[D(A)\omega]_{x,i}{}^a,$$

by (3.15) and (A.31).

With this result, we may write the one-plaquette action (A.35), to order $\varepsilon^{3/2}$, as

$$S_{x,i} = \varepsilon\,(4n)^{-1} \sum_b \{ \psi^b{}_a(V_{x,i})\, v_{x,i}{}^a + [D(V_{x,i})\omega]_{x,i}{}^a \}^2 , \qquad (A.38a)$$

or, in matrix notation

$$S_{x,i} = \varepsilon\,(4n)^{-1} \,(\, \psi(V)\, v + D(V)\omega,\ \psi(V)\, v + D(V)\omega\,) , \qquad (A.38a)$$

where $V = 2^{-1}(A+B)$, and v is a vector with independent components $v^\alpha$, that represents a generic configuration defined on horizontal links of the lattice. This expression has a simple geometric interpretation. Recall that

$$ds^2 = \sum_b [\, \psi^b{}_c(A)\, dA^c\, ]^2 . \qquad (A.39)$$

is the line element at A of the SU(n) group that is invariant under left and right multiplication. With $B-A = \varepsilon^{1/2}v$, and $g = \exp(\varepsilon^{1/2}\omega)$, we see that to order $\varepsilon^{3/2}$, the one-plaquette action is the line element of the difference of the horizontal links B-A, plus the gauge covariant derivative of the gauge-transformation g defined by the vertical links, evaluated at the mean $V = 2^{-1}(B+A)$. This expression is a lattice analog of the continuum expression $F_{0i}{}^2 = (\partial_0 A_i - D_i(A)A_0)^2$.

Step 4. Integration over vertical links, and lattice form of Gauss's Law

We now integrate out the variables $\omega_x$ associated with the vertical links by Gaussian integration, with the result

$$\int d\omega \exp[- \varepsilon^{-1}\Sigma_{x,i} S_{x,i}]$$

$$= \det{}^{-1/2}[D^*(V)D(V)] \exp[ - (4n)^{-1} (v, G(V)v) ]. \quad (A.40)$$

The matrix G(V) defines a metric tensor on transverse link configurations which is the lattice analog of the gauge-invariant metric introduced in continuum theory by Babelon and Viallet [19]. Its explicit expression is

$$G(V) \equiv P^{tr} \psi^*(V) [ 1 - D (D^*D)^{-1}D^* ] \psi(V) P^{tr}, \quad (A.41)$$

where $D \equiv D(V)$ and $D^* \equiv D^*(V)$. Here $P^{tr}$ is the projector onto transverse configurations

$$P^{tr} \equiv 1 - \nabla(\nabla^*\nabla)^{-1}\nabla^*. \quad (A.42)$$

It appears on the left and right of G, because in (A.41), v is transverse, $\nabla^*v = 0$, so the metric tensor G(V) in (A.40) is defined only on transverse configurations. Each factor in (A.41) is a matrix that acts on link variables. The middle factor in G(V),

$$P^G \equiv 1 - D (D^*D)^{-1}D^* \quad (A.43)$$

is the projector onto the subspace that satisfies Gauss's law, $D^*(V)P^G = 0$, and in consequence, the electric field operator defined below will satisfy the lattice form of Gauss's law exactly.

Step 5. Integration over horizontal links

With this result, we obtain from (A.21) to order $\varepsilon$,

$$(I_1^{\alpha}, I_2^{\alpha,\beta}) = N \int dv \, \det\psi(B) \, \det M(B) \, \det{}^{-1/2}[D^*(V)D(V)]$$

$$\times \exp[ - (4n)^{-1} (v, G(V)v) ] \, (\varepsilon^{1/2}v^{\alpha}, 2^{-1}\varepsilon v^{\alpha}v^{\beta}) \,, \quad \text{(A.44)},$$

where $B = A + \varepsilon^{1/2}v$, and $V = A + 2^{-1}\varepsilon^{1/2}v$, $\det\psi(A) \equiv \prod_{x,i}\det\psi(A_{x,i})$. Because $I_2$ contains an explicit factor of $\varepsilon$, we may set $\varepsilon$ to zero everywhere else. This gives

$$I_2^{\alpha,\beta} = I_0^{(0)} \, n\varepsilon \, [G^{-1}(A)]^{\alpha,\beta}, \qquad \text{(A.45)}$$

where

$$I_0^{(0)} = C \det\psi(A) \det[D^*(A)\nabla] \det{}^{-1/2}[D^*(A)D(A)] \det{}^{-1/2}G(A), \qquad \text{(A.46)}$$

$M(A) = D^*(A)\nabla$, and C is a constant which we shall determine shortly. Here $I_0^{(0)}$ is the value of $I_0$, to zeroth order in $\varepsilon$, which one obtains if $I_0$ is evaluated by direct integration, as we have done for $I_2$. However by (A.16) we have, $I_0 = 1$, which is exact for all $\varepsilon$ and A, so also

$$I_0^{(0)} = 1. \qquad (A.47)$$

This determines the value of the constant C, which is found by setting $A = 0$, namely $C = \det^{-1/2}(\nabla^*\nabla)$. It also yields the value of detG(A), namely

$$\det^{1/2}G(A) = \det\psi(A)\ \det(D^*(A)\nabla)$$

$$\times \det^{-1/2}[\mathcal{D}^*(A)\mathcal{D}(A)]\ \det^{-1/2}(\nabla^*\nabla). \qquad (A.48)$$

We thus obtain for $I_2^{\alpha,\beta}$,

$$I_2^{\alpha,\beta} = n\varepsilon\ [G^{-1}(A)]^{\alpha,\beta}. \qquad (A.49)$$

There remains to evaluate $I_1^{\alpha}$. Expression (A.44) for $I_1^{\alpha}$ contains an explicit factor of $\varepsilon^{1/2}$ in the right hand factor, so it is sufficient to evaluate the first factor to order $\varepsilon^{1/2}$. Therefore in the first factor, where $B = A+\varepsilon^{1/2}v$, and $V = A+2^{-1}\varepsilon^{1/2}v$, we may set

$$\det\psi(B)\ \det M(B) = (\ 1 + \varepsilon^{1/2}\ v^{\beta}\ \partial/\partial A^{\beta}\ )\ [\det\psi(A)\ \det M(A)] \qquad (A.50a)$$

and

$$\det^{-1/2}[\mathcal{D}^*(V)\mathcal{D}(V)]\ \exp[\ -\ (4n)^{-1}\ (v,\ G(V)v)\ ]$$

$$= (1 + 2^{-1}\varepsilon^{1/2}\, v^\beta\, \partial/\partial A^\beta)\ \{\det^{-1/2}[D^*(A)D(A)]$$

$$\times \exp[\,-(4n)^{-1}\,(v,\, G(A)v)]\}. \qquad \text{(A.50b)}$$

This gives, with

$$\sigma(A) \equiv \det\psi(A)\, \det M(A) \qquad \text{(A.51)}$$

$$I_1{}^\alpha = 2n\varepsilon\, \sigma^{-1}(A)\, [G^{-1}(A)]^{\alpha,\beta}\, \partial\sigma(A)/\partial A^\beta$$

$$+\, n\varepsilon\, \sigma(A)\, \partial/\partial A^\beta\{\, \sigma^{-1}(A)\, [G^{-1}(A)]^{\alpha,\beta}\, \}$$

$$I_1{}^\alpha = n\varepsilon\, \sigma^{-1}(A)\, \partial/\partial A^\beta\, \{\sigma(A)\, [G^{-1}(A)]^{\alpha,\beta}\}, \qquad \text{(A.52)}$$

and we have used (A.48)

Upon substituting these results into (A.8), we obtain finally, with $\varepsilon = g_0{}^2\, a_0\, (2na)^{-1}$,

$$(T'\Psi)(A) = \Psi(A^{tr}) - a_0\, K\Psi(A), \qquad \text{(A.53)}$$

where K is the kinetic energy operator

$$K = -\, g_0{}^2\, (2a)^{-1}\, \sigma^{-1}(A)\, \partial/\partial A^\alpha\, \sigma(A)\, [G^{-1}(A)]^{\alpha,\beta}\, \partial/\partial A^\beta. \qquad \text{(A.54)}$$

Step 6. Lattice color-electric field operator and lattice color-Coulomb operator

The explicit form of $G^{-1}(A)$ is given by

$$G^{-1} = [\ 1 - D(\nabla^* D)^{-1}\nabla^*]\ J^*\ J\ [\ 1 - \nabla(D^*\nabla)^{-1}D^*], \qquad \text{(A.55)}$$

where $D = D(A)$, and $\nabla^* D = D^* \nabla = M$, where $M = M(A)$ is the Faddeev-Popov operator. Here $J = J(A)$ is a matrix which is diagonal on links that is inverse to the Maurer-Cartan form, namely on each link it satisfies

$$J_a{}^b(A_{x,i})\ \psi^a{}_c(A_{x,i}) = \delta_c{}^b. \qquad \text{(A.56)}$$

The $J_a{}^b(A_{x,i})$ satisfy the Lie differential equations of the SU(n) group,

$$[J_a, J_b] = f_{abc} J_c, \qquad \text{(A.57)}$$

where

$$J_a \equiv J_a{}^b(A_{x,i})\partial/\partial A_{x,i}{}^b. \qquad \text{(A.58)}$$

To prove (A.55), we shall verify that

$$G\ Q = P^{tr}, \qquad \text{(A.59)}$$

where $P^{tr}$ is the unit operator on the space of transverse link variables. Here Q is the expression on the right hand side of (A.55), and G is given in (A.41). We obviously have $\nabla^* Q = Q\nabla = 0$, so

$$P^{tr} Q = Q, \tag{A.60}$$

and moreover

$$\psi P^{tr} Q = \psi Q = [ 1 - \mathcal{D}(\nabla^* D)^{-1}\nabla^* J^*] J [ 1 - \nabla(D^*\nabla)^{-1}D^*],$$

where $\mathcal{D} = \psi D$. This gives, as required,

$$G Q = P^{tr} \psi^* [ 1 - \mathcal{D} (\mathcal{D}^*\mathcal{D})^{-1}\mathcal{D}^* ] [ 1 - \mathcal{D}(\nabla^* D)^{-1}\nabla^* J^* ]$$

$$\times\ J [ 1 - \nabla(D^*\nabla)^{-1}D^*]$$

$$= P^{tr} \psi^* [ 1 - \mathcal{D} (\mathcal{D}^*\mathcal{D})^{-1}\mathcal{D}^* ] J [ 1 - \nabla(D^*\nabla)^{-1}D^*]$$

$$= P^{tr} \psi^* [ J - \mathcal{D} (\mathcal{D}^*\mathcal{D})^{-1}D^* ] [ 1 - \nabla(D^*\nabla)^{-1}D^*]$$

$$= P^{tr} \psi^* J [ 1 - \nabla(D^*\nabla)^{-1}D^*] = P^{tr} [ 1 - \nabla(D^*\nabla)^{-1}D^*] = P^{tr},$$

where we have used $\mathcal{D}^* J = D^*$.

## Appendix B: Explicit form of Transformed Hamiltonian

After transformation (5.15), the hamiltonian is given by the quadratic form

$$(\Phi', H_{eff}' \Psi') = (\Phi', K' \Psi') + (\Phi', (V_m + \gamma_0 G) \Psi'), \qquad \text{(B.1)}$$

where $V_m$ is the magnetic field energy of (4.16), G is given in (5.1), and the transformed kinetic energy operator is given by

$$(\Psi', K'\Phi') = \int dA^{tr} [g_0^2(2a)^{-1}] \ \Sigma_{x,i} (E_{x,i,a}\Phi' - (2\sigma)^{-1} [E_{x,i,a}, \sigma]\Phi')^*$$

$$\times \ (E_{x,i,a}\Psi' - (2\sigma)^{-1} [ E_{x,i,a}, \sigma] \Psi'), \qquad \text{(B.2a)}$$

where the operator

$$E_{x,a,i} = E_{x,a,i}^{\alpha}(A^{tr}) \, i \, \partial/\partial A^{tr,\alpha} \qquad \text{(B.2b)}$$

is given in (4.17) and (4.18). With

$$E_{x,a,i}^{\dagger} \equiv i \, \partial/\partial A^{tr,\alpha} E_{x,a,i}^{\alpha}(A^{tr}), \qquad \text{(B.2c)}$$

the transformed hamiltonian operator acts according to

$$H_{eff}' = K'\Psi' + (V_m + \gamma_0 G) \Psi', \qquad \text{(B.3)}$$

where

$$K'\Psi' = [g_0^2(2a)^{-1}] \sum_{x,i} \{E_{x,i}^{a\dagger} E_{x,i}^{a} - [E_{x,i,a}^{\alpha} (2\sigma)^{-1}\partial\sigma/\partial A^{tr,\alpha}]^2$$

$$+ \partial [E_{x,i,a}^{\alpha} E_{x,i,a}^{\beta} (2\sigma)^{-1}\partial\sigma/\partial A^{tr,\beta}] /\partial A^{tr,\alpha} \} \Psi'. \quad (B.4)$$

The last two terms represent an effective potential energy.

## Appendix C. Properties of the Faddeev-Popov Propagator

We write the Faddeev-Popov matrix as

$$M = M_0 + M_1, \quad (C.1)$$

where

$$M_0 \equiv \nabla^*\nabla \quad (C.2)$$

$$M_1 \equiv g_0\, B^*\nabla = g_0\, \nabla^* B \quad (C.3)$$

and the operator B is defined by

$$g_0\, B \equiv D - \nabla, \quad (C.4)$$

where D is the lattice gauge-covariant derivative of sect. 3. We have

$$M^{-1} = M_0^{-1} - M_0^{-1} M_1 M_0^{-1} + M_0^{-1} M_1 M^{-1} M_1 M_0^{-1}. \quad (C.5)$$

By the definition of D

$$(D_i^{ac}\omega^c)_x = G_{x,i}^{ac}\nabla_i\omega^c + g_0 f^{abc}A_{x,i}^b a_i\omega_x^c, \qquad (C.6)$$

where $a_i\omega_x = 2^{-1}[\omega(x+e_i) + \omega(x)]$, the matrix $M^{-1}$ has the particular structure

$$M^{-1} = M_0^{-1} + M_0^{-1}\,\nabla^*F\nabla M_0^{-1} - g_0\, M_0^{-1}\,\nabla^*A\, M_0^{-1}. \qquad (C.7)$$

Here the action of the operator A is defined by the last term of (C.6), and F is defined by

$$F_{x,i,a;y,j,b} \equiv g_0^2\, B^{(x)}{}_i^{a,c}\, B^{(y)}{}_j^{b,d}\, (M^{-1})_{x,c;y,d}$$

$$- (G_{x,i}^{a,b} - \delta^{a,b})\, \delta_{x,y}\;\; \delta_{i,j}. \qquad (C.8)$$

The explicit expression for $G_{x,i}^{a,b}$ is given in (3.7).

Consider the related matrix E defined by

$$E_{x,i,a;y,j,b} \equiv E1_{x,i,a;y,j,b} + E2_{x,i,a;y,j,b} \qquad (C.9)$$

where

$$E1_{x,i,a;y,j,b} \equiv D^{(x)}{}_i^{a,c}\, D^{(y)}{}_j^{b,d}\, (M^{-1})_{x,c;y,d}. \qquad (C.10)$$

$$E2_{x,i,a;y,j,b} \equiv - G_{x,i}^{a,b}\, \delta_{x,y}\;\; \delta_{i,j}. \qquad (C.11)$$

Because $D = g_0 B + \nabla$, the matrices E and F are related by

$$F_{x,i,a;y,j,b}{}^{TT} = [\, E_{x,i,a;y,j,b} + \delta_{x,y}\, \delta_{i,j}\, \delta^{a,b} \,]^{TT} \qquad (C.12)$$

where the superscript TT means the part that is transverse on both indices. The matrix E is related to the horizon function G by

$$G = \Sigma_{x,y,i,a}\, E_{x,i,a;y,i,a}. \qquad (C.13)$$

The propagator C, defined in (6.1) may be written

$$C = M_0^{-1} + M_0^{-1}\, \nabla^* \langle F \rangle\, \nabla M_0^{-1}, \qquad (C.14)$$

where we have used $\langle A \rangle = 0$, which follows from global color or lattice symmetry. In terms of its fourier transform (6.2), this reads

$$C_\theta = \Lambda_\theta^{-1} + \Lambda_\theta^{-2}\, q_{\theta,i} K_{\theta,i,j}\, q_{\theta,j} \qquad (C.15)$$

where

$$K_{\theta,i,j}\, \delta^{a,b} \equiv V^{-1}\, \Sigma_{x,y}\, \exp[-i\theta\cdot(x + 2^{-1}e_i - y - 2^{-1}e_j)]\, \langle F_{x,i,a;y,j,b} \rangle. \qquad (C.16)$$

We also define

$$L_{\theta,i,j}\, \delta^{a,b} \equiv V^{-1} \sum_{x,y} \exp[-i\theta\cdot(x + 2^{-1}e_i - y - 2^{-1}e_j)]\, \langle E_{x,i,a;y,j,b}\rangle, \tag{C.17}$$

which has the decomposition

$$L_{\theta,i,j} = L_{1\theta,i,j} + L_{2\theta,i,j}. \tag{C.18}$$

corresponding to the decomposition $E = E_1 + E_2$. By (C.12), $K_{\theta,i,j}$ and $L_{\theta,i,j}$ are related by

$$K_{\theta,i,j}{}^{TT} = [\, L_{\theta,i,j} + \delta_{i,j} \,]^{TT}. \tag{C.19}$$

We now establish a number of properties of $L_{\theta,i,j}$. By (C.11) and by lattice and global color symmetry we have

$$L_{2\theta,i,j} = -\, b\, \delta_{i,j}, \tag{C.20}$$

where

$$b\, \delta^{a,c} \equiv \langle G_{x,i}{}^{a,c}\rangle = \delta^{a,c} \langle n^{-1} \operatorname{Re} \operatorname{tr} U_{x,i}\rangle, \tag{C.21}$$

by (3.7). The matrix $E_1$ satisfies

$$\nabla^* E_{1x,a;y,j,b} = D^{(y)}{}_j{}^{b,d}\, \delta_{x,y}\, \delta_{a,d}, \tag{C.22}$$

and thus

$$\Sigma_i \, q_{\theta,i} \, L_{1\theta,i,j} = \; b \, q_{\theta,j}, \qquad \text{(C.23)}$$

where we have again used $\langle A \rangle = 0$. This gives

$$L_{1\theta,i,j} = b \, \delta_{i,j} + S_{\theta,i,j}{}^{TT}, \qquad \text{(C.24)}$$

where $S_{\theta,i,j}{}^{TT}$ is transverse on both indices. We have

$$L_{\theta,i,j} = S_{\theta,i,j}{}^{TT}, \qquad \text{(C.25)}$$

so $L_{\theta,i,j}$ is transverse on both indices. From (C.13), the horizon condition, $\langle G \rangle = 0$, is equivalent to the statement

$$\Sigma_i \, L_{\theta,i,i}|_{\theta=0} = 0. \qquad \text{(C.26)}$$

So far the volume of the lattice is finite. We now take the infinite volume limit, so $\theta_i$ becomes a continuous angular variable, and we write $L_{i,j}(\theta)$, and similarly for other $\theta$-dependent quantities. We assume that global color symmetry and cubic symmetry of the lattice are preserved in the infinite-volume limit. We also assume that the leading terms at $\theta_i = 0$ have the invariance of the full (3-dimensional) Euclidean group and not just the symmetries of the cubic lattice group. This means that $L_{i,j}(\theta)$ and $K_{i,j}(\theta)$ are Euclidean-invariant functions of $q(\theta)$ at small $\theta$. Then (C.25) implies that in the neighborhood of $\theta_i = 0$, the tensor $L_{i,j}(q)$ is of the form

$$L_{i,j}(q) \approx a(q^2) \, ( \, \delta_{i,j} - q_i \, q_j \, /q^2 \, ), \qquad \text{(C.27)}$$

because this is the only transverse tensor with the stated symmetries. By eq. (C.26), the horizon condition is the statement

$$a(0) = 0, \tag{C.28}$$

and we have $L_{i,j}(0)= 0$. According to (C.19), this is equivalent to the statement

$$[\ K(0)_{i,j} - \delta_{i,j}\ ]^{TT} = 0, \tag{C.29}$$

so the horizon condition has been expressed as a property of $C(\theta)$.

We wish to express $C(\theta)$ in terms of its self-energy. Let the matrix $\Sigma(\theta)$ with elements $\Sigma_{i,j}(\theta)$ be defined by

$$\Sigma(\theta) \equiv K(\theta)\ [\ 1 + P(\theta)\ K(\theta)\ ]^{-1}, \tag{C.30}$$

where $P(\theta)$ is the projector onto the longitudinal subspace,

$$P_{i,j}(\theta) \equiv q_i(\theta)\ q_j(\theta)\ /\ q^2(\theta). \tag{C.31}$$

It has the expansion

$$\Sigma_{i,j}(\theta) = K_{i,j}(\theta) - K_{i,k}(\theta)\ q_k(\theta)\ [q^2(\theta)]^{-1}\ q_l(\theta)\ K_{l,j}(\theta) + \dots\ . \tag{C.32}$$

Equation (C.30) may be inverted to give

$$K(\theta) \equiv \Sigma(\theta)\ [\ 1 - P(\theta)\ \Sigma(\theta)\ ]^{-1}, \qquad \text{(C.33)}$$

If this is substituted into eq. (C.15) for $C(\theta)$ one obtains (6.3).

The expansion (C.32) shows that the transverse parts of $K_{i,j}(\theta)$ and $\Sigma_{i,j}(\theta)$ are equal at $\theta = 0$,

$$[\Sigma_{i,j}(0)]^{TT} = [K_{i,j}(0)]^{TT}\ , \qquad \text{(C.34)}$$

and we conclude that the horizon condition in the form (C.29) is equivalent to (6.6).

In expression (C.32) for $\Sigma_{\theta,i,j}$, the leading singularity of $K_{i,j}(\theta)$ at $\theta = 0$, is canceled by the remaining terms. For $K_{i,j}(\theta)$, we use (C.16) and (C.8), and obtain from (C.32)

$$\Sigma_{\theta,i,j} = [(n^2-1)V]^{-1} \sum_{x,y} \exp[-i\theta\cdot(x + 2^{-1}e_i - y - 2^{-1}e_j)]$$

$$\times\ g_0^2\ \langle B^{(x)}{}_i{}^{a,c}\ B^{(y)}{}_j{}^{a,d}\ (M^{-1})_{x,c;y,d}\rangle + \ldots\ . \qquad \text{(C.35)}$$

This gives

$$\Sigma_{\theta,i,j} \equiv [(n^2-1)V^3]^{-1} \sum_{x,y,\theta',\theta''} \exp[-i\theta\cdot(x + 2^{-1}e_i - y - 2^{-1}e_j)]$$

$$\times\, g_0^2 \, \langle (M^{-1})_{\theta',c;\theta'',d} \, B^{(x)}{}_i{}^{a,c} \, B^{(y)}{}_j{}^{a,d} \rangle \, \exp[i(\theta'\cdot x - \theta''\cdot y)] + \ldots , \qquad (C.36)$$

where $(M^{-1})_{\theta',c;\theta'',d}$ is defined in (6.11). We use (C.4) and (C.6) for $B^{(x)}{}_i{}^{a,c}$, and write explicitly only the leading term in $g_0$, to obtain (6.10).

## Appendix D. Comparison with the Renormalization-group and Perturbative Calculations

The energy E(r) of a pair of external quarks at separation r is a gauge-invariant quantity of the form

$$E(r) = (4\pi r)^{-1} \, g_{rg}{}^2(r) \, \langle Q_1{}^a \, Q_2{}^a \rangle, \qquad (D.1)$$

where $g_{rg}(r)$ is the renormalized running coupling constant of QCD. At small separation r it is given by [20],

$$g_{rg}{}^{-2}(r) = b_{rg} \ln[\, (r\Lambda_{QCD})^{-1} \,] , \qquad r << \Lambda_{QCD}{}^{-1} \qquad (D.2)$$

$$b_{rg} = (24\pi^2)^{-1} \, (11n - 2n_f) , \qquad (D.3)$$

where $n_f$ is the number of quark flavors, and $\Lambda_{QCD}$ is a hadronic mass. Expressions (8.21) and (D.2) exhibit the same analytic dependence on r, but the numerical coefficients $b_c$ and $b_{rg}$ are different.

To understand this difference, it is sufficient to consider the perturbative calculation of E(r) up to one-loop order, because the renormalization-group coefficient $b_{rg}$ is determined by the one-loop perturbative calculation. The perturbative calculation to one-loop in the Coulomb gauge is presented in [20], with the result

$$E(r) = (4\pi r)^{-1} g_p^2(r) \langle Q_1^a Q_2^a \rangle, \qquad (D.4)$$

where

$$g_p^2(r) = g_0^2 + g_0^4 b_{rg} \ln(r\Lambda) + ... \qquad (D.5)$$

and $\Lambda$ is an ultraviolet cut-off. The $g_0^2$ term is the classical Coulomb energy, and the $g_0^4$ term is the one-loop contribution, and $b_{rg}$ is the renormalization-group coefficient. In the Coulomb gauge, it results from the sum of two contributions [20],

$$b_{rg} = \delta + \delta' \qquad (D.6)$$

where

$$\delta = (2\pi^2)^{-1} n \qquad (D.7)$$

$$\delta' = - (24\pi^2)^{-1} ( n + 2n_f ), \qquad (D.8)$$

whose physical origin we now explain. Both terms result from the part of the Coulomb hamiltonian given by

$$H_1 = 2^{-1} \sum_{x,y} \rho_x^a \; [ g_0 (M^{-1})^{ab}( - \nabla^2 )^{-1} g_0 (M^{-1})^{bc}]_{x,y} \rho_y^c, \tag{D.9}$$

where $\rho = \rho_{dyn} + \rho_{ext}$. The contribution $\delta'$ results from the second-order quantum-mechanical perturbative formula

$$E^{(2)} = - \sum_n | \langle 0| H_1|n\rangle |^2 (E_n - E_0)^{-1}, \tag{D.10}$$

where to order $g_0^4$, one makes the replacements $M^{-1}(-\nabla^2)M^{-1} \to (-\nabla^2)^{-1}$. This contribution is purely negative and corresponds to the familiar vacuum polarization or screening of QED. Thus $\delta'$ should not contribute to the static color-Coulomb coefficient $b_c$, because it comes from a distortion of the wave-function caused by the presence of external charges, and this is not part of what we have called the static color-Coulomb interaction.

The contribution $\delta$, which is positive and represents anti-screening, comes from treating $H_1$ in first-order quantum-mechanical perturbation theory

$$E^{(1)} = \langle 0| H_1|0\rangle, \tag{D.11}$$

evaluated to order $g_0^4$. We have

$$E^{(1)} = 2^{-1} \sum_{x,y} \langle 0| \rho_{ext,x}^a \, [(g_0M^{-1})^{ab}( - \nabla^2 ) (g_0M^{-1})^{bc}]_{x,y} \, \rho_{ext,y}^c |0\rangle.$$

$$E^{(1)} = \langle 0|\, [\, (g_0 M^{-1})^{ab} ( -\nabla^2) (g_0 M^{-1})^{bc}\, ]_{r,0}\, |0\rangle\, \langle Q_1{}^a\, Q_2{}^c \rangle. \qquad \text{(D.12)}$$

The Faddeev-Popov operator, $M = -\nabla^2 - g_0\lambda$, where $\lambda \equiv A\cdot\nabla$, has the expansion in powers of $g_0$

$$M^{-1}( -\nabla^2)\, M^{-1} = ( -\nabla^2)^{-1} + 2\, ( -\nabla^2)^{-1}\, g_0\lambda\, ( -\nabla^2)^{-1}$$

$$+ 3\, ( -\nabla^2)^{-1}\, g_0\lambda\, ( -\nabla^2)^{-1}\, g_0\lambda\, ( -\nabla^2)^{-1} + \ldots \qquad \text{(D.13)}$$

The $g_0{}^4$ contribution comes from the last term, with coefficient 3, which gives the contribution $\delta$ to $b_{rg}$.

We may evaluate separately the contribution to $E^{(1)}$ that comes from the vacuum intermediate state, and the part that comes from excited intermediate states,

$$\langle 0|\, M^{-1}( -\nabla^2)\, M^{-1}\, |0\rangle = \langle 0|\, M^{-1}|0\rangle\, ( -\nabla^2)\, \langle 0|\, M^{-1}|0\rangle$$

$$+ \sum_{n\neq 0} \langle 0|\, M^{-1}\, |n\rangle( -\nabla^2)\, \langle n|\, M^{-1}\, |0\rangle. \qquad \text{(D.14)}$$

The two parts contribute additively to $\delta$ and $b_{rg}$,

$$b_{rg} = \delta + \delta' = \delta_v + \delta_e + \delta', \qquad \text{(D.15)}$$

Only the first term in (D.14) corresponds to what we have called the color-Coulomb interaction energy, because it is obtained by the mean-field substitution $M^{-1} \to \langle M^{-1}\rangle$, so consistency requires $b_c = \delta_v$.

The perturbative expansion of the vacuum-intermediate-state contribution to (D.14) is given by

$$\langle M^{-1}\rangle ( -\nabla^2) \langle M^{-1} \rangle = ( - \nabla^2)^{-1}$$

$$+ 2 \langle ( - \nabla^2)^{-1} g_0\lambda ( - \nabla^2)^{-1} g_0\lambda ( - \nabla^2)^{-1} \rangle + \ldots \quad \text{(D.16)}$$

The last term has coefficient 2 as compared to the coefficient 3 in (D.13), and so, by (D.7),

$$\delta_v = (2/3)\, \delta = (3\pi^2)^{-1}\, n, \quad \text{(D.17)}$$

and indeed, by comparison with (8.21) one has, as required

$$b_c = \delta_v. \quad \text{(D.18)}$$

We conclude that the lowest order calculation in the present approach gives directly the renormalization-group result for the static color-Coulombic interaction energy. A renormalized running coupling constant $g_c(r)$ has been obtained in the present approach without any renormalization having been done. The mass $\mu$ corresponds to $\Lambda_{QCD}$.

Qualitatively, $b_c$ is the largest contribution to $b_{rg}$, being 8/11 of the gluonic contribution to $b_{rg}$. For $n_f = 6$ flavors of quarks and for $n = 3$ of SU(3), the ratio is $b_c/b_{rg} = 8/7$. Although the corrections to $b_c$ are numerically significant (and calculable), they do not produce a qualitative change.

## Appendix E: Lower bound on quark-anti-quark potential

Consider the hamiltonian $H_{0,tot}$, eq. (7.34), which describes an external quark and antiquark at lattice sites 0 and r in interaction with dynamical gluons. (Dynamical quarks would screen the interaction of interest.) The color-charge density operator of the external quarks is given by

$$\rho_{ext,x}{}^{a} = i\,( t^{(1)a}\, \delta_{x,0} + t^{(2)a*}\, \delta_{x,r} ), \qquad \text{(E.1)}$$

where $t^{(i)a}$ and $t^{(i)a*}$ are anti-hermitian generators of SU(n) in the fundamental representation and its complex conjugate representation respectively. The hamiltonian (7.34) has the form

$$H_{0,tot} = H_{glue} + H_{cross\text{-}term} + H_{ext}, \qquad \text{(E.2a)}$$

where $H_{glue} = H_0$, eq. (7.20),

$$H_{ext} = [2\,(2\pi)^3]^{-1} \int d^3q\; \rho_{ext}{}^{a}(q)\; v(q)\; \rho_{ext}{}^{a}(q)\;, \qquad \text{(E.2b)}$$

and $H_{cross\text{-}term}$ contains the terms $\rho_{ext}\, v\, \rho_{glue}$.

As trial wave-function we choose

$$\Psi_{\alpha,\beta} = n^{-1/2}\, \delta_{\alpha,\beta}\, \Psi_0, \qquad (E.3)$$

where $t^{(1)}$ and $t^{(2)}$ act on the first and second indices respectively, and $\Psi_0$ is the vacuum or ground state of $H_{glue}$, so $H_{glue}\Psi_0 = 0$. The quantum-mechanical energy E(r) of the ground state of the system with the external quarks satisfies the bound $E(r) \leq F(r)$, where

$$F(r) \equiv (\Psi, H_{0,tot}\, \Psi). \qquad (E.4)$$

For the wave-function (E.3), $H_{cross\text{-}term}$ gives vanishing contribution to F(r), and we obtain

$$F(r) \equiv (\Psi, H_{ext}\, \Psi). \qquad (E.5)$$

The color-charge density operator (E.1) acts on the wave-function (E.3) according to

$$(\rho_{ext,x}{}^{a}\, \Psi)_{\alpha,\beta} = i\, n^{-1/2}\, (t)^{a}{}_{\beta,\alpha}\, (\, \delta_{x,0} - \delta_{x,r}\, )\, \Psi_0, \qquad (E.6)$$

because $[(t^a)_{\alpha,\beta}]^* = -\,(t^a)_{\beta,\alpha}$, which reads, in momentum space,

$$(\rho_{ext}{}^{a}(q)\, \Psi)_{\alpha,\beta} = i\, n^{-1/2}\, (t)^{a}{}_{\beta,\alpha}\, [\, 1 - \exp(iq\cdot r)\, ]\, \Psi_0\, . \qquad (E.7)$$

We obtain for F(r),

$$F(r) = [2(2\pi)^3]^{-1}\, n^{-1}\, (n^2-1) \int d^3q\, v(q)\, [\, 1 - \cos(q\cdot r)\, ], \qquad (E.8)$$

where we have used $tr(-t^a t^a) = 2^{-1}(n^2-1)$. For large $|r|$, we may replace $v(q)$ by its asymptotic form at low momentum $v^{as}(q)$. This gives for the energy of a quark-antiquark pair at large separation r,

$$F^{as}(r) = [2(2\pi)^3]^{-1} n^{-1} (n^2-1) \int d^3q\, v^{as}(q) [ 1 - \cos(q\cdot r) ]. \quad (E.9)$$

For the asymptotic form which is a power law, $v^{as}(q) = C/q^p$ we obtain

$$F^{as}(r) = (2\pi)^{-2} n^{-1} (n^2-1) C r^{p-3} \int_0^\infty dx\, x^{2-p} [ 1 - x^{-1}\sin x ],$$

which converges for $3 < p < 5$, with the result

$$F^{as}(r) = (2\pi)^{-2} n^{-1} (n^2-1) C r^{p-3}$$

$$\times [(p-2)(p-3)(p-4)]^{-1} \sin(\pi p/2)\, \Gamma(5-p) . \quad (E.10)$$

If $v^{as}(q)$ has the form (9.9), with $p = 14/3$, one has

$$F^{as}(r) \sim r^{5/3}. \quad (E.11)$$

This expression for $F^{as}$ and the upper bound $E^{as}(r) \leq F^{as}(r)$ are consistent with a linearly rising quark-anti-quark energy $E^{as}(r) = Kr$.

**Figure Captions**

Fig. 1. Plots of $f(\alpha) \equiv N^{-1}(\langle G \rangle_\alpha + \alpha^{-1})$ against $\alpha$, see eqs. (2.22), (2.24).

(i) The number of degrees of freedom N is finite.

(ii) As $N \to \infty$, $\alpha_{cr} \to 0$. This is case (a), the perturbative phase.

(iii) As $N \to \infty$, $\alpha_{cr}$ remains finite. This is case (b), the horizon phase.